\documentclass[
prb,twocolumn,preprintnumbers,amsmath,amssymb]{revtex4}

\usepackage{graphicx}
\usepackage{amssymb}
\usepackage{amsmath}
\usepackage{dsfont}
\usepackage{bm}
\usepackage{mathrsfs}
\usepackage{times}

\usepackage{feynmf}
\usepackage{amsfonts}

\usepackage{array}
\newcolumntype{C}[1]{>{\centering\arraybackslash$}p{#1}<{$}}

\begin{document}
\title{
Reflection and transmission in nonlocal susceptibility models with multiple resonances
}

\author{R.\ J.\ Churchill}

\email{rc313@exeter.ac.uk}

\author{T.\ G.\ Philbin}

\email{t.g.philbin@exeter.ac.uk}

\affiliation{Physics and Astronomy Department, University of Exeter,
Stocker Road, Exeter EX4 4QL, United Kingdom}

\begin{abstract}
We consider a semi-infinite dielectric with multiple spatially dispersive resonances in the susceptibility. 
The effect of the boundary is described by an arbitrary reflection coefficient for polarization waves in the material at the surface, with specific values corresponding to various additional boundary conditions (ABCs) for Maxwell's equations.
We derive exact expressions for the electromagnetic reflection and transmission coefficients and present the results for a variety of materials with multiple exciton bands. We find an improved single-band approximation for heavy/light exciton bands and extend our model to exciton dispersion relations with linear $k$ terms which occur in uniaxial crystals. Finally, we calculate the spectral energy density of thermal and zero-point radiation for a variety of multi-resonance models and ABCs.
\end{abstract}

\maketitle

\section{Introduction}

The susceptibility describing the material response to an applied electromagnetic field contains both temporal and spatial dispersion\cite{LLcm,rukhadze1961}.
As a result of the latter, the induced polarization at any point in the medium depends on the applied field in the region surrounding that point.
This behavior can be due to a range of excitations within the material, such as phonons or excitons, each with their own associated resonance in the susceptibility.
While this nonlocal response is often negligible in comparison to the frequency dependence there are cases where it can play a significant role, such as 
metallic nanostructures\cite{raz11,wie12,pen12,tos15,sch16}, 
radiative heat transfer\cite{sin15a,sin15b}, 
spontaneous emission\cite{pur46,dre68,bar98,Novotny,hor14}, 
spectral energy density\cite{churchill2016b} and 
Casimir self-forces\cite{hor14}.
Spatial dispersion is also important in semiconductors, where the complex electronic band structure can lead to many excitations\cite{honerlage1984}, each with their own nonlocal response.
In this paper, we continue our previous work on nonlocal response\cite{churchill2016b} by extending results on reflection and transmission at planar boundaries to the case where the medium has multiple spatially-dispersive resonances.

The susceptibility is typically expressed as a sum of resonances.
Nonlocal response is included as a $k$ dependence in the model parameters, but is usually limited to a $k$-dependent resonant frequency.
Hopfield and Thomas\cite{hopfieldthomas} proposed the following model, based on the properties of semiconductors, but it can also be derived from a simple classical model\cite{chu16}:
\begin{align}
\chi({\bm k},\omega)
=
\chi_0
+
\sum_{m=1}^M
\frac{\omega_{pm}^2}{\omega^2_{Tm}(k)-\omega^2-i\gamma_m\omega},
\label{eq:susc}
\end{align}
where $\omega_T(k)$ is the resonant frequency, $\gamma$ quantifies the absorption and $\omega_p$ is the oscillator strength.
The term $\chi_0$ collects contributions from other resonances and acts as a background susceptibility.
For the sake of simplicity, the parabolic dispersion
\begin{align}
\hbar\omega_{T}(k)=\hbar\omega_{T}+\frac{\hbar^2k^2}{2m_{\rm ex}},
\qquad
\omega^2_{T}(k)
\approx\omega^2_{T}+Dk^2
\label{eq:parabolic}
\end{align}
was used\cite{hopfieldthomas} to describe the exciton bands in the medium, where $m_{\rm ex}$ is the exciton mass and $D=\hbar\omega_T/m_{\rm ex}$

The difficulties involved in the calculation of electromagnetic reflection and transmission coefficients for nonlocal media are well known.
While there is only one transmitted wave for a local medium, the nonlocal medium has several transmitted waves due to the $k$ dependence in (\ref{eq:susc})\cite{rukhadze1961}.
The Maxwell boundary conditions are no longer sufficient to solve for the unknown amplitudes of the multiple transmitted waves.
Historically, this need for extra information was resolved with the introduction of Additional Boundary Conditions (ABCs) on the polarization ${\bm P}$ associated with the resonances in (\ref{eq:susc}).
Various authors\cite{agarwal1971a,agarwal1971b,agarwal1972,birman1972,agarwal1973,maradudin1973,birman1974,mills1974,foley1975,bishop1976,ting1975,kliewer1968,kliewer1971,kliewer1975,ruppin1981,rimbey1974,rimbey1975,rimbey1976,rimbey1977,rimbey1978,pekar1958a,pekar1958b,pekar1958c,pekar1959}
 have proposed different ABCs under certain assumptions that suit different types of material.
The Pekar ABC, where ${\bm P}$ vanishes at the boundary, is the simplest and most commonly used.

The majority of work on the subject has focused on susceptibilities with a single, isolated resonance.
In this case, Halevi and Fuchs\cite{Halevi} have derived reflection coefficients for a generalized ABC model, containing all the previously suggested ABCs.
In a previous paper\cite{churchill2016b} we have adapted this model to the tensor case with different transverse and longitudinal susceptibilities.

In general, the susceptibility of real materials is far more complex than simple isolated resonances\cite{honerlage1984}.
In exciton bands, for example, there can be multiple closely spaced bands, degenerate bands\cite{honerlage1984,kane1975,altarelli1977} and more complex $k$ dependence\cite{koteles1979,koteles1980,Fiorini1980}.
While some authors\cite{lagois1977,lagois1981,venghaus1978,sell1973,sermage1981,mahan1964,halevi1985,halevi1987} have considered multi-resonance systems, they are typically limited to a maximum of two resonances and a specific ABC.

The first aim of this paper is to extend the Halevi and Fuchs\cite{Halevi} generalized ABC model to a multi-resonance susceptibility and derive expressions for the reflection and transmission coefficients.
This derivation is first applied to a system with simple parabolic exciton bands and then to bands that are degenerate at $k=0$, where we find improved parameters for the single band approximation.

The second aim is to modify the derivation further to include alternate wave vector dependences, specifically the case where the dispersion in (\ref{eq:parabolic}) contains a $\pm k$ term.
This behavior is known as linear splitting and is typically found in uniaxial crystals such as Wurtzite.
This case has been previously calculated\cite{mahan1964,halevi1985,halevi1987}, but only for a specific ABC and orientation of the crystal axis.
We will show that linear $k$ splitting can easily be incorporated into the multi-resonance model and that the orientation of the crystal axis has significant effects on the result.

Finally, we will use the derived electromagnetic reflection coefficients to calculate the spectral energy density of thermal and zero-point radiation outside the spatially dispersive medium.
The results for various multi-resonance systems are calculated and compared to those in our previous paper on the isolated resonance\cite{churchill2016b}.

The assumptions made in the following derivation are discussed in greater detail in our previous paper\cite{churchill2016b}.
In summary, we consider a smooth boundary that does not contain any features such as slits or other nontrivial structures and is sufficiently far from any other boundaries such that multiple reflections can be ignored.
In addition we do not include any quantum mechanical effects not directly encoded in the macroscopic susceptibility

The paper is organized as follows.
In Sec. \ref{sec:theory} we present the spatially dispersive susceptibility model for a half-infinite dielectric with a multi-resonance  permittivity and derive the field equations.
In Sec. \ref{sec:p-polarization} and \ref{sec:s-polarization} we derive the general expressions for the reflection and transmission coefficients for $p$ and $s$ polarized light and present the results for parabolic exciton bands in Sec. \ref{sec:results}.
In Sec. \ref{sec:linear} we extend our derivation to a uniaxial crystal by including a linear splitting term and in Sec. \ref{sec:utot} we calculate the zero-point and thermal spectral energy density.

\section{Theory}
\label{sec:theory}

\subsection{Infinite Medium}
We first consider an infinite, homogeneous, spatially-dispersive dielectric with the susceptibility (\ref{eq:susc}).
The electric field ${\bm E}$ and polarization field ${\bm P}$ satisfy the wave equation:
\begin{align}
{\bm \nabla}
\times
{\bm \nabla}
\times
{\bm E}({\bm r},\omega)
-
\frac{\omega^2}{c^2}
{\bm E}({\bm r},\omega)
=
\frac{\omega^2}{c^2}
{\bm P}({\bm r},\omega),
\label{eq:wave_equation}
\end{align}
where the polarization field is given by
\begin{align}
P_i({\bm r},\omega)=
\int  d^3{\bm r}^\prime
\chi({\bm r}-{\bm r}^\prime,\omega)
E_i({\bm r^\prime},\omega).
\label{eq:infinite_polarization}
\end{align}
In general, the spatially dispersive susceptibility $\chi$ is a tensor\cite{LLcm}, but here we consider a scalar.
Using the Fourier transformation
\begin{align}
P_i({\bm r},\omega)
=
\frac{1}{(2\pi)^3}
\int
d^3{\bm k}
P_i({\bm k},\omega)
e^{i {\bm k} \cdot {\bm r}}
\end{align}
we have
\begin{align}
P_i({\bm k},\omega)
=
\sum_j
\chi({\bm k},\omega)
E_i({\bm k},\omega).
\end{align}
The wave equation (\ref{eq:wave_equation}) has solutions for ${\bm E}$ when the frequency and wave vector satisfy the dispersion relation\cite{rukhadze1961}
\begin{align}
(\omega/c)^2\left[1+\chi({\bm k},\omega)\right]=k^2,
\label{eq:transverse_disp_rel}
\end{align}
for transverse waves with ${\bm E}\cdot{\bm k}=0$ or
\begin{align}
1+\chi({\bm k},\omega)=0,
\label{eq:longitudinal_disp_rel}
\end{align}
for longitudinal waves with ${\bm E}\times{\bm k}=0$.
With the field dependence $\textrm{exp}(ik_zz)$ we restrict ourselves to Im$[k_z]>0$, leading to $M+1$ transverse and $M$ longitudinal waves for the susceptibility in (\ref{eq:susc}) and the parabolic dispersion in (\ref{eq:parabolic}).

\subsection{Half-Infinite Medium}

\begin{figure}[!htb]\centering
{\includegraphics[width=\linewidth]{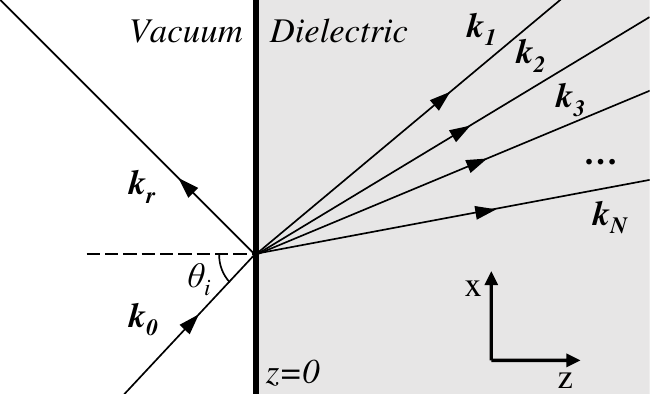}}
\caption{
Schematic of the model.
The $z<0$ vacuum half-space contains the incident (${\bm k_0}$) and reflected (${\bm k_r}$) wave.
The angle of incidence is $\theta_i$.
The $z>0$ spatially dispersive dielectric half-space contains $M+1$ transverse (${\bm k_1}$, ${\bm k_2}\dots{\bm k_{M+1}}$) and $M$ longitudinal (${\bm k_{M+2}}\dots{\bm k_{N}}$) transmitted waves.
The coordinate system is chosen such that the $xz$-plane coincides with the plane of incidence and $k_y=0$.
}
\label{fig:model}
\end{figure}

We now consider the half-infinite dielectric occupying the $z>0$ region as shown in Fig. \ref{fig:model}.
The vacuum contains the incident wave  (${\bm E_0}$) and reflected wave (${\bm E_r}$) with wave-vectors $\bm{k}_0$ and $\bm{k}_r$ ($k_0=k_r=\omega/c$).
Inside the dielectric there are $N=2M+1$ transmitted waves (${\bm E_n}$) with the corresponding wave vectors ${\bm k_n}$.
The coordinate system has been chosen such that the $xz$-plane coincides with the plane of incidence, with $k_{nx}=K$, $k_{ny}=0$ and $k_{nz}=q_n$.

The bulk susceptibility (\ref{eq:susc}) in this co-ordinate system can be rewritten as:
\begin{align}
\chi(K,0,q)=
&\chi_0
+\sum^M_{m=1}
\chi_m(K,0,q),
\end{align}
where:
\begin{align}
\chi_m(K,0,q)=
\frac{\omega^2_{p m}/D_m}{q^2-\Gamma_m^2(K)}
\end{align}
and:
\begin{align}
\Gamma_m^2(K)
=
\frac{\omega^2-\omega_{T m}^2+i\gamma_m\omega}{D_m}-K^2.
\end{align}
With the presence of the boundary at $z=0$, the polarization field now depends on a position-dependent susceptibility $\chi^\prime_i$ ($i\in\{x,y,z\}$)\cite{Halevi}. After a Fourier transformation in the $xy$ plane:
\begin{align}
P_i(K,0,z)=
\int_{0}^{\infty} dz^\prime
\tilde{\chi}^\prime_{i}(K,0,z,z^\prime)
E_i(K,0,z^\prime).
\label{eq:polarization_half_infinite}
\end{align}
We subsequently omit $K$ dependence for notational simplicity.
We assume that each resonance in the half-infinite medium can be expressed in terms of the bulk susceptibility in the same manner as Halevi and Fuchs\cite{Halevi}:
\begin{equation}
\tilde{\chi}^\prime_{m i}(z,z^\prime)=
\left\{
\begin{aligned}
&\tilde{\chi}_m(z-z^\prime)+U_{m i}\tilde{\chi}_m(z+z^\prime)&\text{if } z,z^\prime > 0, \\
&0&\text{otherwise}
\end{aligned}
\right.
\label{eq:susc_U}
\end{equation}
and the overall susceptibility is given by:
\begin{equation}
\tilde{\chi}^\prime_{i}(z,z^\prime)=
\left\{
\begin{aligned}
&\chi_0\delta(z-z^\prime)+\sum_{m}^M\tilde{\chi}^\prime_{m i}(z,z^\prime)&\text{if } z,z^\prime > 0, \\
&0&\text{otherwise}.
\end{aligned}
\right.
\label{eq:susc_U_full}
\end{equation}

The first term in (\ref{eq:susc_U}) is the position-independent nonlocal bulk response.
The second describes a polarization wave propagating from $z^\prime$ to the surface before reflecting and continuing to $z$.
The reflection amplitude coefficient $U_i$ is (in general) complex and frequency dependent, with $|U_i|=1$ implying elastic reflection.
Halevi \& Fuchs\cite{Halevi} demonstrated that specific values of $U_i$ correspond to certain ABC's, shown in Table \ref{tab:abc}.
Each ABC was developed for a particular type of medium or excitation.
For example, Pekar\cite{pekar1958a,pekar1958b,pekar1958c,pekar1959}
 and Rimbey-Mahan\cite{rimbey1974,rimbey1975,rimbey1976,rimbey1977,rimbey1978} 
  were developed for Frenkel (tight-binding) excitons, Ting \emph{et al.}\cite{ting1975}
   for Wannier-Mott (weak-binding) excitons, Fuchs-Kleiwer\cite{ting1975,kliewer1968,kliewer1971,kliewer1975,ruppin1981} for metals and Agarwal \emph{et al.}\cite{agarwal1971a,agarwal1971b,agarwal1972,birman1972,agarwal1973,maradudin1973,birman1974,mills1974,foley1975,bishop1976} for the general case where surface effects can be ignored.

\begin{table}[!htb]\centering
\caption{\label{tab:abc}List of ABC's}
\begin{ruledtabular}
\begin{tabular}{llll}
 & $U_{x}$ & $U_{y}$ & $U_{z}$\\
 \hline
Agarwal \emph{et al}.\cite{agarwal1971a,agarwal1971b,agarwal1972,birman1972,agarwal1973,maradudin1973,birman1974,mills1974,foley1975,bishop1976}
 & \phantom{-}0 & \phantom{-}0 & \phantom{-}0\\
Ting \emph{et al}.\cite{ting1975}& \phantom{-}1 & \phantom{-}1 & \phantom{-}1\\
Fuchs-Kliewer\cite{ting1975,kliewer1968,kliewer1971,kliewer1975,ruppin1981} & \phantom{-}1 & \phantom{-}1 & -1\\
Rimbey-Mahan\cite{rimbey1974,rimbey1975,rimbey1976,rimbey1977,rimbey1978} & -1 & -1 & \phantom{-}1\\
Pekar\cite{pekar1958a,pekar1958b,pekar1958c,pekar1959} & -1 & -1 & -1\\
\end{tabular}
\end{ruledtabular}
\end{table}

Substituting  (\ref{eq:susc_U_full}) into (\ref{eq:polarization_half_infinite}) gives:
\begin{align}
P_i(z)=&\chi_0E_i(z)+
\frac{1}{2\pi}
\int_{-\infty}^{\infty} dq
\int_{0}^{\infty} dz^\prime
\sum_{m=1}^M
\nonumber\\&
\bigg[
e^{iq(z-z^\prime)}
+
U_{m i}
e^{iq(z+z^\prime)}
\bigg]
\chi_m(q)
E_i(z^\prime)
,
\quad
z>0.
\label{eq:polarization_integral}
\end{align}
At this point we introduce an ansatz for the ${\bm E}$ field inside the medium\cite{Halevi} --- a linear combination of $N=2M+1$ plane waves from (\ref{eq:transverse_disp_rel}) and (\ref{eq:longitudinal_disp_rel}):
\begin{align}
E_i(z)
=
\sum_{n=1}^{N}
E_i^{(n)}
e^{iq_nz},
\label{eq:E_ansatz}
\end{align}
where $n=1$ to $M+1$ are transverse waves and $n=M+2$ to $N$ are longitudinal waves.

After substitution of the ansatz (\ref{eq:E_ansatz}) into (\ref{eq:polarization_integral}) and evaluating the $z^\prime$ integral, we find
\begin{align}
P_i(z)=&\chi_0E_i(z)+
\frac{i}{2\pi}
\int_{-\infty}^{\infty} dq
e^{iqz}
\sum_{m=1}^M
\sum_{n=1}^N
\nonumber\\&
\bigg[
\frac{1}{q_n-q}
+
\frac{U_{m i}}{q_n+q}
\bigg]
\chi_m(q)
E_i^{(n)}
,
\quad
z>0.
\label{eq:polarization_integral_new}
\end{align}
The $q$ integral is evaluated by performing a contour integration in the upper half-plane.
This encloses the poles at $q=q_n$ and $\Gamma_m$, giving
\begin{align}
P_i(z)&=
\sum_{n=1}^{N}\chi(q_n)E_i^{(n)}e^{iq_nz}
\nonumber\\&-\frac{\omega_{p 1}^2}{2D_1\Gamma_1}
\sum_{n=1}^{N}
\left(\frac{1}{q_n-\Gamma_1}+\frac{U_{1 i}}{q_n+\Gamma_1}\right)
E_i^{(n)}e^{i\Gamma_1z}
\nonumber\\&-\frac{\omega_{p 2}^2}{2D_2\Gamma_2}
\sum_{n=1}^{N}
\left(\frac{1}{q_n-\Gamma_2}+\frac{U_{2 i}}{q_n+\Gamma_2}\right)
E_i^{(n)}e^{i\Gamma_2z}
\nonumber\\&-
\dots
\nonumber\\&-\frac{\omega_{p M}^2}{2D_M\Gamma_M}
\sum_{n=1}^{N}
\left(\frac{1}{q_n-\Gamma_M}+\frac{U_{M i}}{q_n+\Gamma_M}\right)
E_i^{(n)}e^{i\Gamma_Mz}.
\label{eq:P_full}
\end{align}
For the wave equation (\ref{eq:wave_equation}) to be valid for all values of $z$, we require each of the right-hand side sums proportional to $\textrm{exp}(i\Gamma_mz)$ in (\ref{eq:P_full}) to equal zero:
\begin{align}
\sum_{n=1}^{N}
\phi_{m i}^{(n)}
E_i^{(n)}
=0,
\qquad
m=1,\dots,M
\label{eq:field_amp_eqs}
\end{align}
where
\begin{align}
\phi_{m i}^{(n)}
=
\left(\frac{1}{q_n-\Gamma_m}+\frac{U_{m i}}{q_n+\Gamma_m}\right).
\end{align}
This leads to a set of $M$ equations of the form (\ref{eq:field_amp_eqs}) for each of the $E_i$ components.

\section{$p$-polarization}
\label{sec:p-polarization}
The field can be decomposed to components with $\bm{E}$ perpendicular to ($s$-polarized) or in  the plane of incidence ($p$-polarized).
For $p$-polarized light $E_y=0$, $E_x\ne0$ and $E_z\ne0$.
Both the transverse and longitudinal waves appear in the medium.
\subsection{Surface Impedance}
The reflection coefficient is calculated using the surface impedance, which for $p$-polarized light is given by:
\begin{align}
Z_p
=
\frac{
E_x(0^+)
}{
H_y(0^+)
}.
\end{align}
(Here ${\bm H}=\mu_0{\bm B}$.) The magnetic field $B_y$ can be expressed in terms of the electric field using $k_0{\bm B}={\bm k}\times{\bm E}$ and (\ref{eq:E_ansatz}): 
\begin{align}
B_y^{(n)}(z)&
=
\frac{1}{k_0}
\left[
q_nE_x^{(n)}
-
KE_z^{(n)}
\right]
e^{iq_nz}
\nonumber\\&
=
\left[
\frac{
q_n
-
K\eta^{(n)}
}{k_0}
\right]
E_x^{(n)}
e^{iq_nz}
\nonumber\\&
=
\tau^{(n)}
E_x^{(n)}
e^{iq_nz},
\end{align}
where we have substituted $E_x$ for $E_z$ using
\begin{align}
E_z^{(n)}=\eta^{(n)}E_x^{(n)},
\label{eq:Ex_Ez}
\end{align}
where $\eta^{(n)}=-K/q_n$ for transverse waves and $\eta^{(n)}=q_n/K$ for longitudinal waves. This leads to:
\begin{equation}
\tau^{(n)}=
\left\{
\begin{aligned}
&\frac{q_n^2+K^2}{q_n k_0},	&\text{transverse waves},	\\
&0,								&\text{longitudinal waves}.
\
\end{aligned}
\right.
\label{eq:tau}
\end{equation}
The surface impedance can now be expressed solely in terms of $E_x$ field amplitude ratios:
\begin{align}
Z_p
=&
\frac{1}{\mu_0}
\frac{
\sum_{n=1}^N
E_x^{(n)}
}{
\sum_{n=1}^N
\tau^{(n)}
E_x^{(n)}
}
\nonumber\\
=&
\frac{1}{\mu_0}
\frac{1
+
\sum_{n=2}^N
\frac{E_x^{(n)}}{E_x^{(1)}}
}{
\tau^{(1)}
+
\sum_{n=2}^{N}
\tau^{(n)}
\frac{E_x^{(n)}}{E_x^{(1)}}
}.
\label{eq:surface_imp}
\end{align}

\subsection{Field Amplitude Ratios}
To proceed any further, we require the $E_x$ field amplitude ratios in (\ref{eq:surface_imp}).
By using (\ref{eq:Ex_Ez}), we can rewrite the $E_z$ equations in (\ref{eq:field_amp_eqs}) in terms of $E_x$:
\begin{align}
\sum_{n=1}^{N}
\left[
\phi_{m x}^{(n)}
\right]
E_x^{(n)}
=0,
\qquad
\sum_{n=1}^{N}
\left[
\eta^{(n)}
\phi_{m z}^{(n)}
\right]
E_x^{(n)}
=0.
\label{eq:Ex_amp_sums}
\end{align}
We now have $2M$ equations relating the $2M+1$ waves inside the medium and have sufficient information to solve for the reflection coefficient.
After dividing by $E^{(1)}_x$ and rearranging we can express (\ref{eq:Ex_amp_sums}) in matrix form. As an example, we present the result for a two-resonance system:
\begin{widetext}
\begin{align}
\begin{pmatrix}
\phi_{1x}^{(2)}				&\phi_{1x}^{(3)}			&\phi_{1x}^{(4)}			&\phi_{1x}^{(5)}	\\
\phi_{2x}^{(2)}				&\phi_{2x}^{(3)}			&\phi_{2x}^{(4)}			&\phi_{2x}^{(5)}	\\
\eta^{(2)}\phi_{1z}^{(2)}	&\eta^{(3)}\phi_{1z}^{(3)}	&\eta^{(4)}\phi_{1z}^{(4)}	&\eta^{(5)}\phi_{1z}^{(5)}	\\
\eta^{(2)}\phi_{2z}^{(2)}	&\eta^{(3)}\phi_{2z}^{(3)}	&\eta^{(4)}\phi_{2z}^{(4)}	&\eta^{(5)}\phi_{2z}^{(5)}	\\
\end{pmatrix}
\begin{pmatrix}
E_x^{(2)}/E_x^{(1)}	\\
E_x^{(3)}/E_x^{(1)}	\\
E_x^{(4)}/E_x^{(1)}	\\
E_x^{(5)}/E_x^{(1)}	\\
\end{pmatrix}
=
-
\begin{pmatrix}
\phi_{1x}^{(1)}				\\
\phi_{2x}^{(1)}				\\
\eta^{(1)}\phi_{1z}^{(1)}	\\
\eta^{(1)}\phi_{2z}^{(1)}	\\
\end{pmatrix}
,
\label{eq:ratio_matrix}
\end{align}
\end{widetext}
where $n=1,2,3$ are transverse waves and $n=4,5$ are longitudinal waves.
By inverting the $2M\times2M$ matrix, we can find the field amplitude ratios.

\subsection{Reflection and Transmission Coefficients}
The $p$-polarization reflection coefficient can be expressed in terms of surface impedance\cite{kliewer1968}
\begin{align}
r_p=
\frac{{ E_r}}{{ E_0}}
=
\frac{
Z_p^{(0)}-Z_p
}{
Z_p^{(0)}+Z_p
},
\label{eq:rp}
\end{align}
where $Z_p$ is given by (\ref{eq:surface_imp}) and  $Z_p^{(0)}=\sqrt{k_0^2-K^2}/\mu_0k_0$ is the vacuum surface impedance.

We can find the transmission coefficients for the $N$ transmitted waves by imposing the continuity of the tangential ${\bm E}$  field across the boundary.
Our choice of coordinate system means we simply equate the $E_x$ components on each side:
\begin{align}
\left[E_0-E_r\right]\cos{\theta_i}
=
\left[\sum_{n=1}^NE^{(n)}_x\right].
\end{align}
This can be expressed in terms of the previously calculated field amplitude ratios using (\ref{eq:rp}) and $\cos{\theta_i}=\sqrt{k_0^2-K^2}/k_0$:
\begin{align}
\frac{\sqrt{k_0^2-K^2}}{k_0}\left[1-r_p\right]E_0
=
\left[
1+\sum_{n=2}^N\frac{E^{(n)}_x}{E^{(1)}_x}
\right]
E^{(1)}_x.
\end{align}
By rewriting:
\begin{align}
E^{(n)}
=
\sqrt{
\left[E^{(n)}_x\right]^2
+
\left[E^{(n)}_z\right]^2
}
=\sqrt{1+\eta^{(n)2}}E_x^{(n)},
\end{align}
 we can derive the transmission coefficient:
\begin{align}
t^{(n)}_p
= \frac{{ E^{(n)}}}{{E_0}}
\label{eq:tp_definition}
\end{align}
for transverse waves:
\begin{align}
t^{(n)}_p
=&
\frac{\sqrt{q_n^2+K^2}}{q_n}
\frac{E^{(n)}_x}{E^{(1)}_x}
\frac{\sqrt{k_0^2-K^2}}{k_0}\frac{\left[1-r_p\right]}{\left[
1+\sum_{n=2}^N\frac{E^{(n)}_x}{E^{(1)}_x}
\right]},
\end{align}
and longitudinal waves: 
\begin{align}
t^{(n)}_p
=&
\frac{\sqrt{q_n^2+K^2}}{K}
\frac{E^{(n)}_x}{E^{(1)}_x}
\frac{\sqrt{k_0^2-K^2}}{k_0}\frac{\left[1-r_p\right]}{\left[
1+\sum_{n=2}^N\frac{E^{(n)}_x}{E^{(1)}_x}
\right]}.
\end{align}

We now have a single method for $r_p$ and $t_p$ in the presence of multiple resonances that can cover a wide frequency range.
This derivation could be extended further to a tensor susceptibility using the method described in our previous paper\cite{churchill2016b}.
In this case the polarization reflection coefficents are a tensor $U_{ij}$ with certain restrictions on the components.

\section{$s$-polarization}
\label{sec:s-polarization}
We now consider the simpler case of $s$-polarized light, where $E_y\ne0$, $E_x=0$ and $E_z=0$.
As ${\bm k_n}$ all lie in the $xz$-plane, this leads to the absence of longitudinal waves in the medium, leaving the $M+1$ transverse waves. 
\subsection{Surface Impedance}
The surface impedance for $s$-polarized light is given by:
\begin{align}
Z_s
=&
-
\frac{
E_y(0^+)
}{
H_x(0^+)
}.
\end{align}
As in (\ref{eq:surface_imp}) we express $H_x$ in terms of $E_y$ and field amplitude ratios:
\begin{align}
Z_s
=&
\frac{1}{\mu_0}
\frac{
k_0
\sum_{n=1}^{M+1}
E_y^{(n)}
}{
\sum_{n=1}^{M+1}
q_n
E_y^{(n)}
}
\nonumber\\
=&
\frac{1}{\mu_0}
k_0
\frac{
1+
\sum_{n=2}^{M+1}
\frac{E_y^{(n)}}{E_y^{(1)}}
}{
q_1
+
\sum_{n=2}^{M+1}
q_n
\frac{E_y^{(n)}}{E_y^{(1)}}
}.
\label{eq:Z_s}
\end{align}
As we only have the $M+1$ transverse waves in the $s$-polarization, the set of $M$ equations from (\ref{eq:field_amp_eqs}) can be rewritten as
\begin{align}
\sum_{n=2}^{M+1}
\left[
\phi_{m y}^{(n)}
\right]
\frac{E_y^{(n)}}{E_y^{(1)}}
&=
-\phi_{m y}^{(1)},
\label{eq:field_amp_y}
\end{align}
which is sufficient to solve for the amplitude ratios.

\subsection{Reflection and Transmission Coefficients}
Using the vacuum surface impedance $Z_s^{(0)}=k_0/\mu_0\sqrt{k_0^2-K^2}$ and (\ref{eq:Z_s}), we can construct the $s$-polarization reflection coefficient:
\begin{align}
r_s
=\frac{{ E_r}}{{ E_0}}
=\frac{
Z_s^{(0)}-Z_s
}{
Z_s^{(0)}+Z_s
}.
\label{eq:rs}
\end{align}

As in the previous section, we impose the continuity of the tangential ${\bm E}$  field across the boundary.
As we only have $E_y$ components, this leads to:
\begin{align}
E_0-E_r
=
\sum_{n=1}^{M+1}E_y^{(n)}.
\end{align}
Using (\ref{eq:rs}) this can be expressed in terms of the field amplitude ratios previously found from (\ref{eq:field_amp_y}):
\begin{align}
E_0\left(1-r_s\right)
=&
E_y^{(1)}
\left(1+\sum_{n=2}^{M+1}\frac{E_y^{(n)}}{E_y^{(1)}}\right),
\end{align}
which leads to:
\begin{align}
t_s^{(n)}
=
\frac{E^{(n)}}{E_0}
=&
\frac{E_y^{(n)}}{E_y^{(1)}}
\frac{1-r_s}{1+\sum_{n=2}^{M+1}\frac{E_y^{(n)}}{E_y^{(1)}}}.
\end{align}

\section{Reflection Coefficient Results}
\label{sec:results}

\begin{figure}[!htb]
{\includegraphics[width=\linewidth]{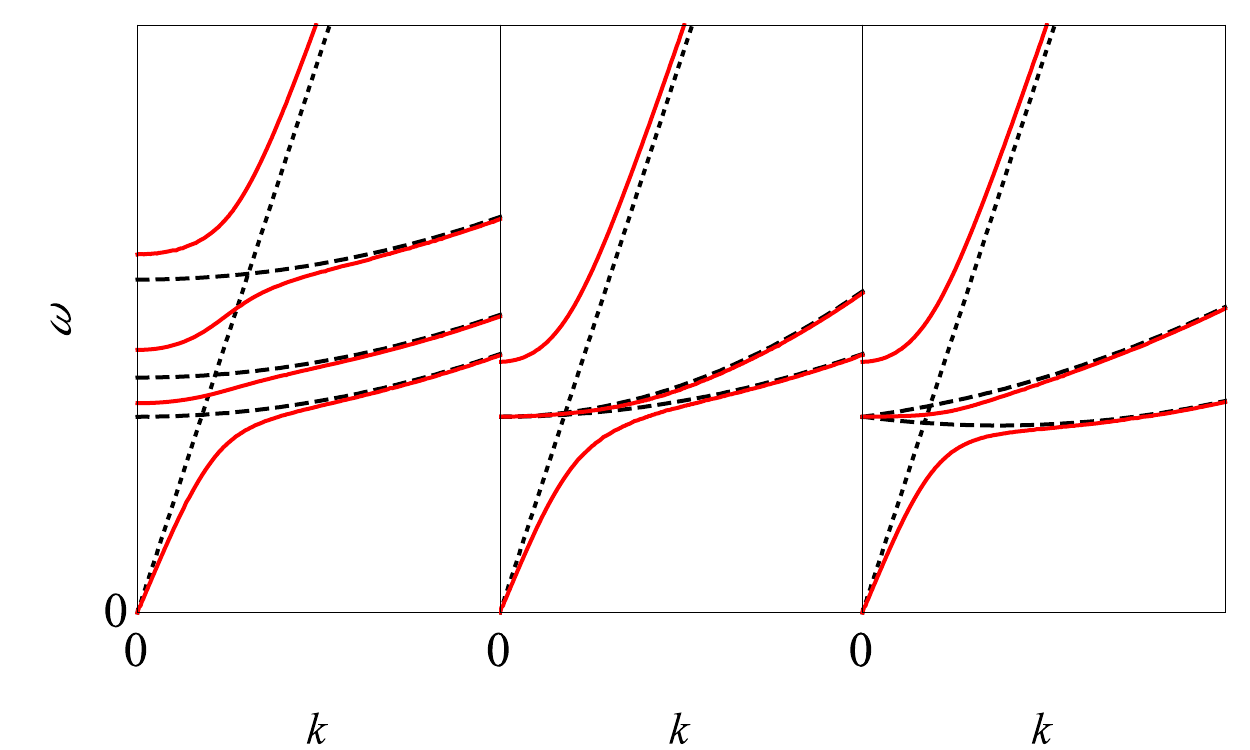}}
\caption{
Exciton band behavior (black dashed) compared to the light line (black dotted) and dispersion relations for transverse ${\bm E}$ waves in an infinite medium (solid red) when $\gamma$ is set to zero.
Examples include multiple parabolic bands (left), heavy/light exciton bands with the same $\omega_T$ but different $k^2$ terms (middle) and bands with the same $\omega_T$ and $k^2$ terms, but a $\pm k$ linear splitting term (right). The exciton bands in last two are degenerate at $k=0$.
}
\label{fig:PAPER[DispRel]}
\end{figure}

We now use the derivations in the previous sections for a number of materials with a variety of exciton band structures.
In Fig. \ref{fig:PAPER[DispRel]} we show some example exciton bands and the corresponding dispersion relations for transverse ${\bm E}$ waves in the absence of damping.

In this section we first consider the simple case of ZnO\cite{lagois1981} with three non-interacting exciton bands and GaAs\cite{sell1973} with two bands that are degenerate at $k=0$.
The model parameters are given in Table \ref{tab:Model_Parameters_1}.
The values of $\omega_p$ are calculated from the measured values of $\omega_L$, which are the solutions of  the dispersion relation for transverse ${\bm E}$ waves at $k=0$ in the absence of damping.
Similarly $D$ is found from the measured exciton mass $m_{\rm ex}$, which is given in units of the rest electron mass $m_{\rm e0}$.

\begin{table}[!htb]\centering
\caption{\label{tab:Model_Parameters_1}List of model parameters}
\begin{ruledtabular}
\begin{tabular}{cccccc}
						&ZnO\cite{lagois1981}	&			& 			&GaAs\cite{sell1973}	\\
m						&1 			&2 			&3 			&1			&2 			\\
						\hline
$\chi_0$					&5.2		&5.2		&5.2		&11.6		&11.6		\\
$\hbar\omega_T$ (eV)	&3.3758		&3.3810		&3.4198		&1.514		&1.514		\\
$\hbar\omega_L$ (eV)	&3.3776		&3.3912		&3.4317		&1.515		&1.515		\\
$\hbar\gamma$ (meV)	&0.7		&0.7		&0.7		&0.05		&0.05		\\
$m_{\rm ex}$ ($m_{\rm ex}$)	&0.87		&0.87		&0.87		&0.183		&0.805		\\
\hline
$\hbar\omega_p$ (eV)	&0.5334		&0.6055		&0.5983		&0.138		&0.138		\\
$D$	 ($10^{11}{\rm m}^2{\rm s}^{-2}$)		&6.82	&6.84	&6.91	&14.55		&3.31		\\
\end{tabular}
\end{ruledtabular}
\end{table}

\subsection{Simple Resonances}

We first consider a three-resonance model for ZnO\cite{lagois1981}, involving the A, B and C excitons which we label $m=1$, $2$ and $3$ respectively.
The exciton bands are of the form in (\ref{eq:parabolic}) and do not interact.

\begin{figure}[!htb]\centering
{\includegraphics[width=\linewidth]{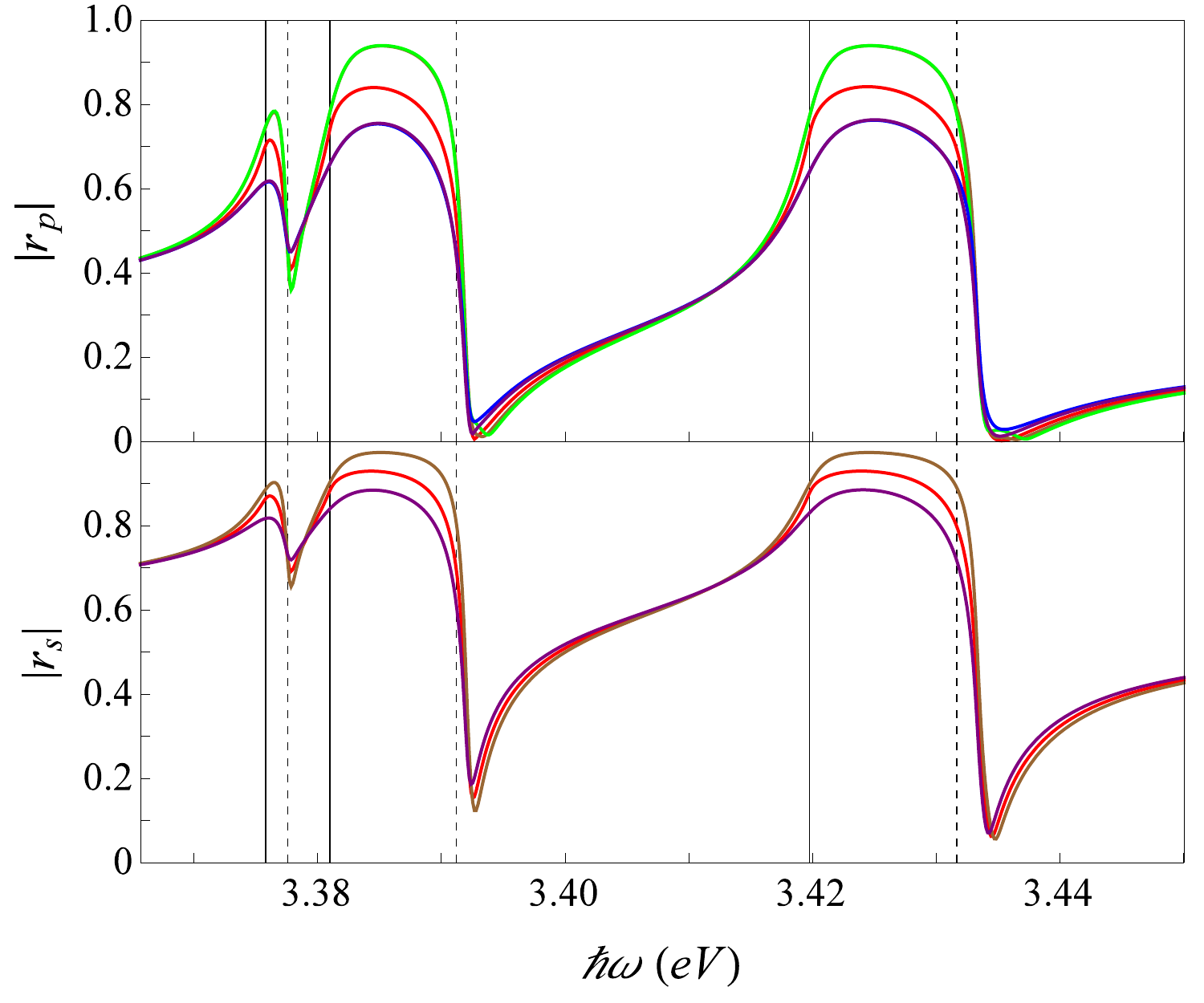}}
\caption{
Reflection coefficients $r_p$ and $r_s$ as a function of $\omega$  for the ZnO 3-exciton model at an incident angle of $60^\circ$.
Vertical lines indicate $\omega_{Tm}$ (solid) and $\omega_{Lm}$ (dashed) values.
Includes Agarwal \emph{et al}. (Red), Ting \emph{et al}. (Brown), Fuchs-Kliewer (Green), Rimbey-Mahan (Blue) and Pekar (Purple) ABC's. 
}
\label{fig:ZnO}
\end{figure}

Figure \ref{fig:ZnO} shows $r_p(\omega)$ and $r_s(\omega)$ for a fixed incident angle.
The peak locations are determined by the $\omega_T$ and $\omega_L$ values, indicated by solid and dashed vertical lines respectively.
The $r_p$ behavior is mostly determined by $U_x$, with $U_x=1$ giving the largest maxima and smallest minima.
In contrast $U_z$ only affects the results at the reflection minima.
The frequency region around $\omega_{T3}$  can be accurately described using the single-resonance model.
This is because the overlap with the other resonances in the susceptibility is very small.
The same is not true of the $m=1$ and $2$ resonances.
Here the proximity of $\omega_{T1}$ and $\omega_{T2}$ lead to the one-resonance models failing, particularly in intermediate frequency region.
The effect of $U_i$ remains the same as the isolated resonance.
However, if there is significant overlap in the resonant peaks (e.g. if the $\omega_T$ values are separated by less that the full-width half-maximum), then $U_x=1$ gives not only the the largest peaks but also the largest value in the intermediate frequency region.

\subsection{Heavy \& Light Excitons}

We now move on to consider an exciton band structure with degeneracy at $k=0$.
Kane\cite{kane1975} showed  that interactions in a medium could lead to the splitting of degenerate exciton bands.
In the case of isotropic valence bands, this can lead to  a ``heavy'' and ``light'' exciton band with parabolic dispersion relations
\begin{align}
\omega^2_{T{\rm h}}(k)
&=\omega^2_{T}+\frac{\hbar\omega_T}{m_{\rm h}}k^2
=\omega^2_{T}+D_{\rm h}k^2,
\\
\omega^2_{T{\rm l}}(k)
&=\omega^2_{T}+\frac{\hbar\omega_T}{m_{\rm l}}k^2
=\omega^2_{T}+D_{\rm l}k^2,
\end{align}
substituted into the susceptibility.

\begin{figure}[!htb]\centering
{\includegraphics[width=\linewidth]{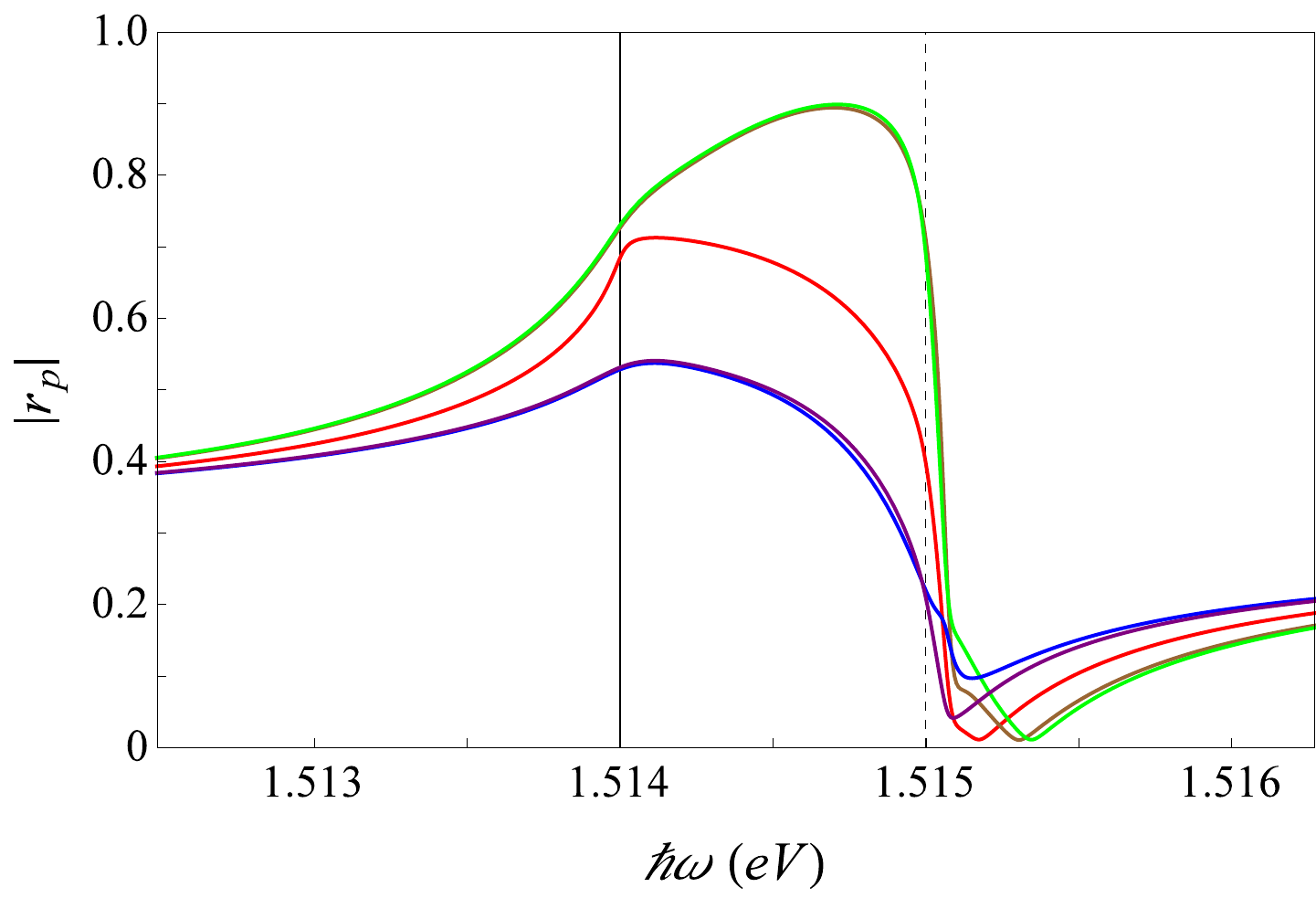}}
\caption{
Reflection coefficient $r_p$ as a function of $\omega$  for the GaAs heavy/light exciton model at an incident angle of $60^\circ$.
Vertical lines indicate $\omega_{T}$ (solid) and $\omega_{L}$ (dashed) values.
Plot styles follow the conventions in  Fig. \ref{fig:ZnO}.
}
\label{fig:GaAs}
\end{figure}
We consider a two-resonance model for GaAs\cite{sell1973} containing only the heavy and light exciton bands, using the parameters in Table \ref{tab:Model_Parameters_1}.
We have slightly simplified the model by assuming isotropic valence bands and using the $\left<100\right>$ exciton masses for all directions.
 The results for $r_p$ are shown in Fig. \ref{fig:GaAs} and display the same basic features as a single resonance model.
The same behavior is seen in the $r_s$ result, which we do not plot.

\begin{figure}[!htb]\centering
{\includegraphics[width=\linewidth]{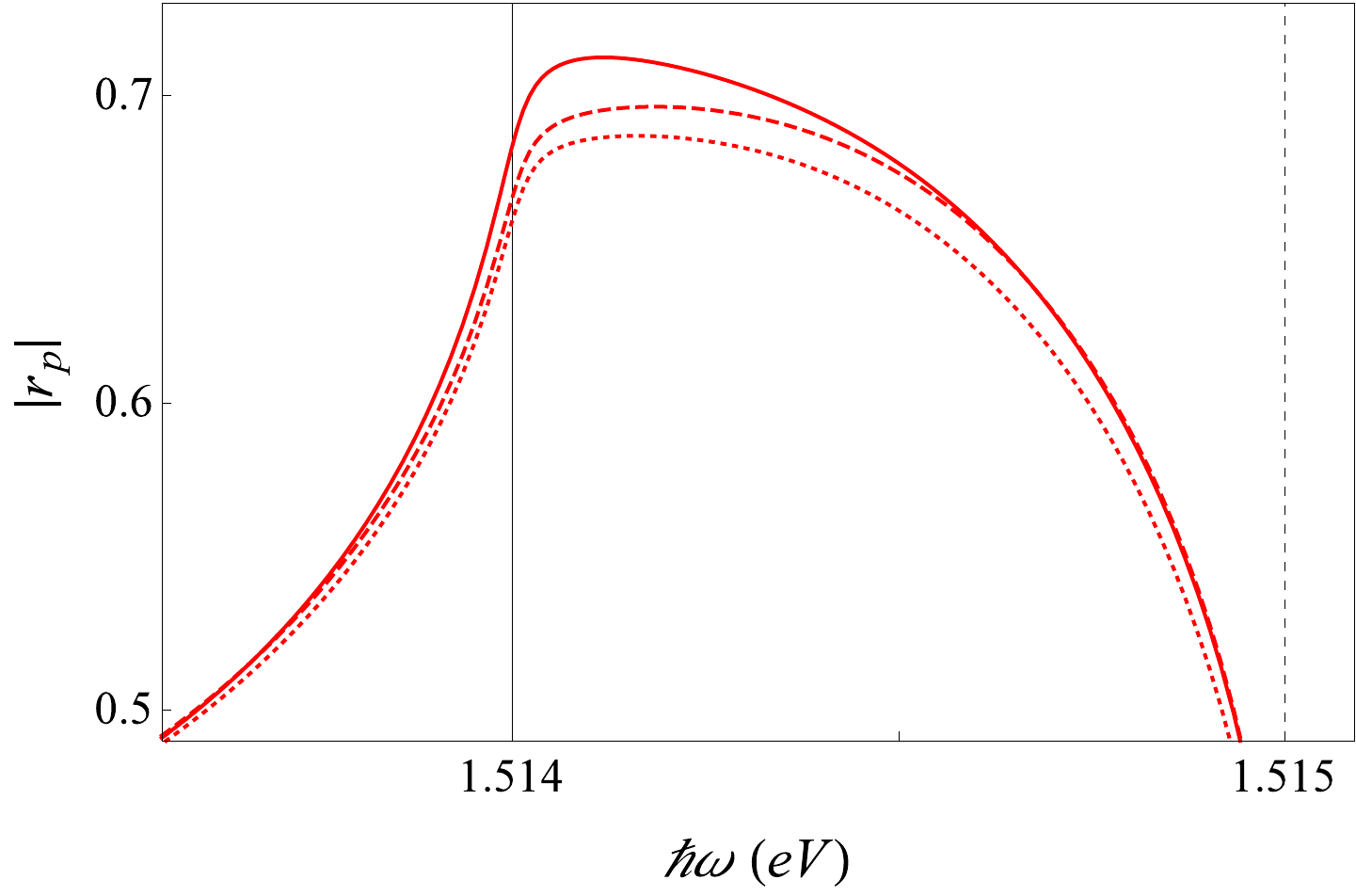}}
\caption{
Detail of $r_p$ using the Agarwal \emph{et al}. ABC for the GaAs heavy/light exciton model (solid line) compared to effective one exciton results.
We find that $D^*$ (dashed) from (\ref{eq:D*}) gives a better fit to the resonant peak than the previously suggested $D_{\rm eff}$ (dotted) from (\ref{eq:Deff}).
}
\label{fig:GaAs_detail}
\end{figure}

Previous work\cite{sell1973} has suggested that the two-resonance system of heavy and light exciton bands can be approximated by a single band with an effective $D_{\rm eff}$ term in the susceptibility, given by
\begin{align}
D_{\rm eff}
=\frac{D_{\rm h}+D_{\rm l}}{2}
\label{eq:Deff}
\end{align}
and multiplying $\omega_p^2$ by a factor of 2.
However, simply taking the average of the nonlocal parameter $D$ as in (\ref{eq:Deff}) does not lead to the best approximation.
Figure \ref{fig:GaAs_detail} compares the heavy/light exciton model to the effective single band result and shows that it underestimates the peak of the heavy/light system. This is true even when $D_{\rm h}$ and $D_{\rm l}$ are close.
Instead, if we re-express the nonlocal term as
\begin{align}
D^*k^2=(\sigma k)^2,
\end{align}
we find that a better fit is given by taking the average value of the coefficient $\sigma$, which leads to:
\begin{align}
\sqrt{D^*}
=\frac{\sqrt{D_{\rm h}}+\sqrt{D_{\rm l}}}{2}
.
\label{eq:D*}
\end{align}
This new value provides an excellent fit when $D_{\rm l}$ and $D_{\rm h}$ have similar values.
For larger differences between $D_{\rm l}$ and $D_{\rm h}$ (such as this model where $D_{\rm l}\approx4D_{\rm h}$), both single-resonance approximations begin to fail as the the two resonance model has a larger peak just above $\omega_T$.
This difference is greatest for $U_x=-1$ and smallest for $U_x=1$.
Despite this, we find that (\ref{eq:D*}) gives a better fit to $r_p$ and $r_s$ than (\ref{eq:Deff}) for all values of $D_{\rm l/h}$ and $U_i$.

\section{Linear $k$ terms}
\label{sec:linear}

So far we have only considered materials with isotropic, parabolic energy bands of the form (\ref{eq:parabolic}).
However the symmetry of the crystal structure can lead to the introduction of additional $k$ terms.
For example, in 1964 Mahan and Hopfield\cite{mahan1964} used a linear $k$ term to explain a shoulder in the reflection spectra of CdS, a uniaxial medium with wurtzite crystal symmetry.
This behavior was only observed when ${\bm E}$ was perpendicular to the crystal axis ${\bm c}$, with the exciton dispersion relation:
\begin{align}
\hbar\omega_{\pm}({\bm k})=\hbar\omega_T
+\frac{\hbar^2k_\perp^2}{2m_{\rm ex\perp}}
+\frac{\hbar^2k_\parallel^2}{2m_{\rm ex\parallel}}
\pm\zeta k_\perp,
\label{eq:linear_split}
\end{align}
where $m_{\rm ex}$ is the exciton mass and $k_\parallel$ and $k_\perp$ are wave vector components parallel and perpendicular  to ${\bm c}$ respectively.
In the susceptibility, (\ref{eq:linear_split}) is approximated to\cite{mahan1964}:
\begin{align}
\omega^2_{\pm}({\bm k})=
\omega^2_T
+D_\perp k_\perp^2
+D_\parallel k_\parallel^2
\pm\xi k_\perp,
\label{eq:linear_split_2}
\end{align}
where $D_{\perp/\parallel}=\hbar\omega_T/m_{\rm ex\perp/\parallel}$ and $\xi=2\omega_T\zeta/\hbar$.
This leads to two resonances: 
\begin{align}
\chi_+({\bm k})+\chi_-({\bm k})=
&\frac{\omega_p^2}{
(\omega^2_T
+D_\perp k_\perp^2
+D_\parallel k_\parallel^2
+\xi k)
-\omega^2-i\gamma\omega
}
\nonumber\\
+&
\frac{\omega_p^2}{
(\omega^2_T
+D_\perp k_\perp^2
+D_\parallel k_\parallel^2
-\xi k)
-\omega^2-i\gamma\omega
},
\end{align}
that can be collected to a single fraction $\chi_{\rm lin}^\perp$, giving
\begin{align}
\chi_{\rm lin}^\perp({\bm k})=
\frac{
2\omega_p^2(\omega^2_T
+D_\perp k_\perp^2
+D_\parallel k_\parallel^2-\omega^2-i\gamma\omega)
}{
(\omega^2_T
+D_\perp k_\perp^2
+D_\parallel k_\parallel^2-\omega^2-i\gamma\omega)^2-\xi^2k_\perp^2
}
\label{eq:chi_perp}
\end{align}
for ${\bm E}\perp{\bm c}$.
When ${\bm E}\parallel{\bm c}$, the exciton bands were found to be degenerate with $\zeta=0$, leading to the resonance:
\begin{align}
\chi_{\rm lin}^{\parallel}({\bm k})=
\frac{2\omega_p^2}{
(\omega^2_T
+D_\perp k_\perp^2
+D_\parallel k_\parallel^2)
-\omega^2-i\gamma\omega
}
\label{eq:chi_par}
\end{align}
The bulk susceptibility of the uniaxial crystal is no longer a scalaer and takes a vector form when ${\bm c}$ is aligned with one of the co-ordinate axes, such as:
\begin{equation}
\chi_{\rm lin}({\bm k})=
\left(
\begin{matrix}
\chi_{\rm lin}^{\perp}({\bm k})	\\
\chi_{\rm lin}^{\perp}({\bm k})	\\
\chi_{\rm lin}^{\parallel}({\bm k})	
 \end{matrix}
 \right)
 \label{eq:chi_vector}
\end{equation}
for ${\bm c}\parallel \hat{{\bm z}}$.
In Tables \ref{tab:chi_y} and \ref{tab:chi_xz} we present the $\chi_{{\rm lin}i}$ components relevant for $s$ and $p$ polarized light when ${\bm c}$ is aligned with each of the co-ordinate axes as defined in Fig. \ref{fig:model}.

Previous work\cite{mahan1964,halevi1985,halevi1987} has only considered such a model for ${\bm c}\parallel \hat{{\bm y}}$ (perpendicular to the incident plane) using the Pekar ABC.
We will now modify the derivation of the previous section to include resonances of the form (\ref{eq:chi_vector}) in a multi-resonance system with arbitrary $U$ values when ${\bm c}$ is aligned with each of the co-ordinate axes.

\begin{table}[!htb]\centering
\caption{List of $\chi_{{\rm lin}y}$ expressions for ${\bm c}$ orientations.}
\begin{ruledtabular}
\begin{tabular}{cccc}
&$\chi_{{\rm lin}y}$		\\
\hline
  \rule[-1.5ex]{0pt}{6ex}
${\bm c}\parallel \hat{{\bm x}}$	
&
$\dfrac{
2\omega_p^2(\omega^2_T
+D_\perp q^2
+D_\parallel K^2-\omega^2-i\gamma\omega)
}{
(\omega^2_T
+D_\perp q^2
+D_\parallel K^2-\omega^2-i\gamma\omega)^2-\xi^2q^2
}$
\\
  \rule[-1.5ex]{0pt}{6ex}
${\bm c}\parallel \hat{{\bm y}}$	
&
$\dfrac{2\omega_p^2}{
(\omega^2_T
+D_\perp (K^2+q^2)
)
-\omega^2-i\gamma\omega
}$
\\
  \rule[-1.5ex]{0pt}{6ex}
${\bm c}\parallel \hat{{\bm z}}$
&
$\dfrac{
2\omega_p^2(\omega^2_T
+D_\perp K^2
+D_\parallel q^2-\omega^2-i\gamma\omega)
}{
(\omega^2_T
+D_\perp K^2
+D_\parallel q^2-\omega^2-i\gamma\omega)^2-\xi^2K^2
}$
\label{tab:chi_y}
\end{tabular}
\end{ruledtabular}
\end{table}

\begin{table*}[!htb]\centering
\caption{List of $\chi_{{\rm lin}x}$ and $\chi_{{\rm lin}z}$ expressions for ${\bm c}$ orientations.}
\begin{ruledtabular}
\begin{tabular}{ccc}
&$\chi_{{\rm lin}x}$ 		&$\chi_{{\rm lin}z}$		\\
\hline
  \rule[-1.5ex]{0pt}{6ex}
${\bm c}\parallel \hat{{\bm x}}$	
&
$\dfrac{2\omega_p^2}{
(\omega^2_T
+D_\perp q^2
+D_\parallel K^2)
-\omega^2-i\gamma\omega
}$
&
$\dfrac{
2\omega_p^2(\omega^2_T
+D_\perp q^2
+D_\parallel K^2-\omega^2-i\gamma\omega)
}{
(\omega^2_T
+D_\perp q^2
+D_\parallel K^2-\omega^2-i\gamma\omega)^2-\xi^2q^2
}$
\\
\hline
  \rule[-1.5ex]{0pt}{6ex}
${\bm c}\parallel \hat{{\bm y}}$	
&
$\dfrac{
2\omega_p^2(\omega^2_T
+D_\perp (K^2+q^2)
-\omega^2-i\gamma\omega)
}{
(\omega^2_T
+D_\perp (K^2+q^2)
-\omega^2-i\gamma\omega)^2-\xi^2(K^2+q^2)
}$
&
$\dfrac{
2\omega_p^2(\omega^2_T
+D_\perp (K^2+q^2)
-\omega^2-i\gamma\omega)
}{
(\omega^2_T
+D_\perp (K^2+q^2)
-\omega^2-i\gamma\omega)^2-\xi^2(K^2+q^2)
}$
\\
\hline
  \rule[-1.5ex]{0pt}{6ex}
${\bm c}\parallel \hat{{\bm z}}$
&
$\dfrac{
2\omega_p^2(\omega^2_T
+D_\perp K^2
+D_\parallel q^2-\omega^2-i\gamma\omega)
}{
(\omega^2_T
+D_\perp K^2
+D_\parallel q^2-\omega^2-i\gamma\omega)^2-\xi^2K^2
}$
&
$\dfrac{2\omega_p^2}{
(\omega^2_T
+D_\perp K^2
+D_\parallel q^2)
-\omega^2-i\gamma\omega
}$
\label{tab:chi_xz}
\end{tabular}
\end{ruledtabular}
\end{table*}

\subsection{Field Amplitude Ratios}

Due to the fact that the bulk components in (\ref{eq:chi_vector}) are no longer equal, the dispersion relations take the form
\begin{align}
k_0^2
\left[
1+\chi_{{\rm lin}y}(q)
\right]
-\left(
K^2+q^2
\right)=0
\label{eq:linear_k_disp_rel_s}
\end{align}
for $s$-polarized light and
\begin{align}
&k_0^2\left[1+\chi_{{\rm lin}x}(q)\right]\left[1+\chi_{{\rm lin}z}(q)\right]
\nonumber\\&
-K^2\left[1+\chi_{{\rm lin}x}(q)\right]
-q^2\left[1+\chi_{{\rm lin}z}(q)\right]
=0
\label{eq:linear_k_disp_rel_p}
\end{align}
for $p$-polarized light, where we have omitted $K$ dependence for notational simplicity.
Unlike the previous section, the results of (\ref{eq:linear_k_disp_rel_s}) are not also solutions of (\ref{eq:linear_k_disp_rel_p}).
We substitute the susceptibility (\ref{eq:chi_perp}-\ref{eq:chi_par}) and the ansatz (\ref{eq:E_ansatz}) into (\ref{eq:polarization_integral}), using the $N$ values of  $q_n$ that satisfy (\ref{eq:linear_k_disp_rel_s}) and (\ref{eq:linear_k_disp_rel_p}).
If the field is aligned with ${\bm c}$, there is no linear splitting and the derivation in the previous section is sufficient to find the amplitude ratios.
If linear splitting is present then $\chi_{{\rm lin}i}$ has two poles with Im$[q]>0$, which we label $\Gamma^{(+)}_{i}$ and $\Gamma^{(-)}_{i}$. 
Evaluating the contour integral in (\ref{eq:polarization_integral_new}) gives
\begin{align}
&
P_i(z)=\sum_{n=1}^{N}
\chi_{{\rm lin}i}(q_n)
E_i^{(n)}e^{iq_nz}
\nonumber\\
&
-F^{(+)}_{i}
\sum_{n=1}^{N}
\left(
\frac{1}{q_n-\Gamma^{(+)}_{i}}
+
\frac{U_{{\rm lin} i}}{q_n+\Gamma^{(+)}_{i}}
\right)
E_i^{(n)}e^{i\Gamma^{(+)}_{i}z}
\nonumber\\
&
-F^{(-)}_{i}
\sum_{n=1}^{N}
\left(
\frac{1}{q_n-\Gamma^{(-)}_{i}}
+
\frac{U_{{\rm lin} i}}{q_n+\Gamma^{(-)}_{i}}
\right)
E_i^{(n)}e^{i\Gamma^{(-)}_{i}z}
,
\end{align}
where $F_{i}^{(\pm)}$ is a simple prefactor and we have used a single value of $U_{{\rm lin}i}$ for both resonances.
The additional terms not proportional to ${\rm exp}(iq_n z)$ lead to the same set of equations as (\ref{eq:field_amp_eqs}), but with different values of $\Gamma_m$:
\begin{align}
\sum_{n=1}^{N}
\left(
\frac{1}{q_n-\Gamma^{(+)}_{i}}
+
\frac{U_{{\rm lin} i}}{q_n+\Gamma^{(+)}_{i}}
\right)
E_i^{(n)}
=&0,
\nonumber\\
\sum_{n=1}^{N}
\left(
\frac{1}{q_n-\Gamma^{(-)}_{i}}
+
\frac{U_{{\rm lin} i}}{q_n+\Gamma^{(-)}_{i}}
\right)
E_i^{(n)}
=&0.
\label{eq:linear_field_amp_ratio}
\end{align}

In summary, the inclusion of linear $k$ terms to the exciton dispersion relation does not significantly affect the derivation of field amplitude ratios presented in the previous section.
The only changes required are to use the appropriate $\Gamma$ values in the field amplitude ratio matrix and the $q$ values that satisfy the dispersion relation.

\subsection{${\bm c}\parallel \hat{{\bm y}}$ Orientation}
The simplest case is to align the crystal axis ${\bm c}$ with $\hat{{\bm y}}$.
This particular orientation has been looked at previously only for the Pekar ABC.
As there is no linear splitting in $\chi_{{\rm lin}y}$ for this orientation, the derivation in the previous section is sufficient to calculate $r_s$ and $t_{s}^{(n)}$ in the $s$-polarization.
The $\chi_{{\rm lin}y}$ term leads to two transverse wave from (\ref{eq:linear_k_disp_rel_s}), if it is the only resonance.

In comparison, $\chi_{{\rm lin}x}$ and $\chi_{{\rm lin}z}$ are equal and contain linear splitting.
Equation (\ref{eq:linear_k_disp_rel_p}) for the $p$-polarization can be simplified to
\begin{align}
\left[1+\chi_{{\rm lin}x}(q)\right]
\left\{
k_0^2\left[1+\chi_{{\rm lin}x}(q)\right]
-
\left(K^2
+q^2\right)
\right\}
=0.
\end{align}
Solutions of the first bracket give two longitudinal waves and the second give three transverse waves, for a total of five if $\chi_{\rm lin}$ is the only resonance.
The transverse waves are no longer the same as those in the $s$-polarization.

As $\chi_{{\rm lin}x}$ and $\chi_{{\rm lin}z}$ contain linear splitting, the integral in (\ref{eq:polarization_integral_new}) leads to two equations of the form (\ref{eq:linear_field_amp_ratio}) for both  $E_x$ and $E_z$.
The $E_z$ equations can be converted to $E_x$ using (\ref{eq:Ex_Ez}) with $\eta^{(n)}=-K/q_n$ for transverse waves and $\eta^{(n)}=q_n/K$ for longitudinal waves as in the previous section.
This gives a total of four $E_x$ equations, the same as the number of transmitted waves added by the resonance, which is sufficient to solve for the reflection coefficient.

\subsection{${\bm c}\parallel \hat{{\bm x}}$ or ${\bm c}\parallel \hat{{\bm z}}$ Orientation}
The two other orientations present additional challenges.
As splitting is present in $\chi_{{\rm lin}y}$, the $s$-polarization now has three transverse waves, and two equations of the form (\ref{eq:linear_field_amp_ratio}) for $E_y$.
There is sufficient information to solve for $r_s$ and $t_s^{(n)}$.

The fact that the expressions for $\chi_{{\rm lin}x}$ and $\chi_{{\rm lin}z}$ in Table \ref{tab:chi_xz} are different means (\ref{eq:linear_k_disp_rel_p}) cannot be simplified and the waves are no longer purely transverse or longitudinal.
The $\chi_{\rm lin}$ resonances give a total of four waves in the absence of other resonances.
The contour integration in (\ref{eq:polarization_integral_new}) leads to a total of three equations of the form  (\ref{eq:linear_field_amp_ratio}) - one from the un-split $\chi_{{\rm lin}i}$ and two from the split $\chi_{{\rm lin}i}$.
Again, we have sufficient information to solve for the field amplitude ratios.

As the waves are no longer purely transverse or longitudinal, care must be taken when converting $E_z$ to $E_x$.
From the wave equation, we find:
\begin{align}
\eta^{(n)}=
-
\frac{1}{Kq_n}
\left\{
k_0^2\left[1+\chi_{{\rm lin}x}(q_n)\right]-q_n^2
\right\}.
\end{align}
Similarly, the relation between $B_y$ and $E_x$ used in (\ref{eq:surface_imp}) is modified from (\ref{eq:tau}) to: 
\begin{align}
\tau^{(n)}=
\left[
\frac{
q_n-K\eta^{(n)}
}{k_0}
\right]
=
\frac{k_0}{q_n}\left[1+\chi_{{\rm lin}x}(q_n)\right].
\label{eq:tau_linear_split}
\end{align}
These expressions and the calculated field amplitude ratios are substituted into (\ref{eq:surface_imp}) and (\ref{eq:Z_s}) to find the surface impedances and the subsequent reflection coefficients.

\begin{figure*}[!htb]
{\includegraphics[width=178mm]{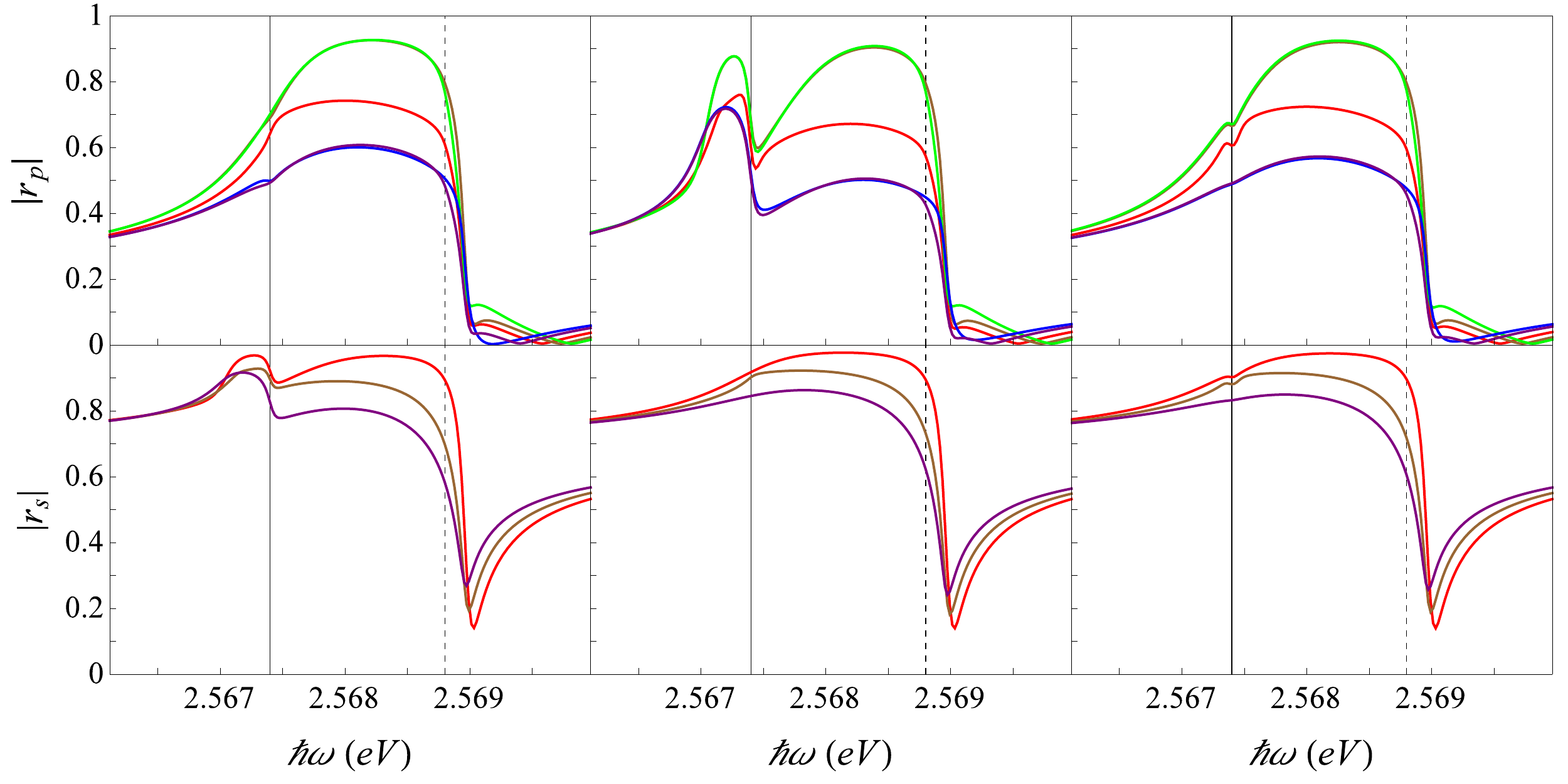}}
\caption{
Reflection coefficients $r_p$ and $r_s$ as a function of $\omega$ at an incident angle of 60$^\circ$ with the crystal axis ${\bm c}$ aligned with $\hat{{\bm x}}$ (left) $\hat{{\bm y}}$ (middle) and $\hat{{\bm z}}$ (right).
Vertical lines indicate $\omega_{T}$ (solid) and $\omega_{L}$ (dashed) values.
Plot styles follow the conventions in  Fig. \ref{fig:ZnO}.
}
\label{fig:CdS[rprs]}
\end{figure*}

\subsection{Results}

As an example, we present the results for CdS\cite{honerlage1984} using the model parameters in Table \ref{tab:Model_Parameters_2}, where $\xi$ in (\ref{eq:linear_split}) has been calculated from the measured $\zeta$ in (\ref{eq:linear_split_2}).
Figure \ref{fig:CdS[rprs]} shows $r_p(\omega)$ and $r_s(\omega)$ at an incident angle of $60^\circ$ for ${\bm c}$ aligned with $\hat{{\bm x}}$, $\hat{{\bm y}}$ and $\hat{{\bm z}}$.
\begin{table}[!htb]\centering
\caption{\label{tab:Model_Parameters_2}List of model parameters}
\begin{ruledtabular}
\begin{tabular}{cccccc}
						&CdS\cite{honerlage1984}	\\
						&						\\
						\hline
$\chi_0$					&6.5 					\\
$\hbar\omega_T$ (eV)	&2.5674					\\
$\hbar\omega_L$ (eV)	&2.5688 				\\
$\hbar\gamma$ (meV)	&0.075					\\
$m_{\rm ex}$ ($m_{\rm ex}$)	&1.3 ($\perp{\bm c}$)	, 1.02 ($\parallel{\bm c}$)	\\
$\zeta$ (eV m)			&5.6$\times10^{-12}	$	\\
\hline
$\hbar\omega_p$ (eV)	&0.164						\\
$D$	 ($10^{11}{\rm m}^2{\rm s}^{-2}$)	& 3.47 ($\perp{\bm c}$), 4.43 ($\parallel{\bm c}$)	\\
$\xi$ (${\rm ms}^{-2}$)					&6.637$\times10^{19}$					\\
\end{tabular}
\end{ruledtabular}
\end{table}

As expected, the ${\bm c}\parallel\hat{{\bm y}}$ result for $r_p$ is significantly different to the others due to linear splitting in both $\chi_{{\rm lin}x}$ and $\chi_{{\rm lin}z}$ components used for $p$ polarized light.
This leads to an additional peak in $r_p$ below $\omega_T$.
While the behavior of the new peak is still mostly determined by the value of $U_x$, the order in which the ABC's appear is different to that of the previous section where $U_x=1$ is the largest and $U_x=-1$ the smallest.
The $r_s$ result for ${\bm c}\parallel\hat{{\bm y}}$ is identical to that of the previous section as there is no splitting in $\chi_{{\rm lin}y}$ for this orientation.

The ${\bm c}\parallel\hat{{\bm x}}$ and ${\bm c}\parallel\hat{{\bm z}}$ cases have splitting in only one component of $\chi_{{\rm lin}i}$ for $p$ polarized light, resulting in an $r_p$ that is closer to the $\xi=0$ result but also displays new features just below $\omega_T$ at a smaller scale.
The ${\bm c}\parallel\hat{{\bm z}}$ results for $r_p$ and $r_s$ are the closest to the $\xi=0$ case as the linear term in Table \ref{tab:chi_xz} only contains $K$.
As $\theta_i$ (and $K$) is decreased, the peak at $\omega_T$ becomes smaller, returning to the $\xi=0$ result for normal incidence.
As in the previous section, the new peak is unaffected by $U_z$ and is larger for $U_x=1$.
In contrast, the ${\bm c}\parallel\hat{{\bm x}}$ case has $q$ in the linear splitting term.
The differences in the $r_p$ result are similar in magnitude to the ${\bm c}\parallel\hat{{\bm z}}$ case, but the new peak at $\omega_T$ is more pronounced for $U_x=-1$ and is now affected by the value of $U_z$.
The effect of the splitting in $q$ is even more pronounced in $r_s$, which displays features similar to $r_p$ in the ${\bm c}\parallel \hat{{\bm y}}$ case.
This is because the linear splitting in $q$ is present in \emph{every} $\chi_{{\rm lin}i}$ term used in their derivation.

In all cases, the difference between $m_\perp$ and $m_\parallel$ has very little effect compared to the choice of ABC.
This agrees with our previous work on the tensor susceptibility\cite{churchill2016b}.

\section{Spectral Energy Density}
\label{sec:utot}

We now focus on the electromagnetic zero-point and thermal radiation at a perpendicular distance $|z|$ from the boundary of the nonlocal medium.
In our previous paper on the one-resonance system\cite{churchill2016b} we found that the inclusion of spatial dispersion removed the unphysical $1/|z|^3$ divergence present in the spectral energy density of the local model\cite{candelas1982,henkel2000}.
We now investigate how the behavior of the materials considered in the previous sections changes due to the presence of multiple resonances.

The average energy density of electromagnetic zero-point and thermal radiation in the vacuum outside a medium is given by\cite{joulain2005}
\begin{align}
\langle U\rangle
=&
\frac{\varepsilon_0}{2}\langle \left|{\bm E}\left({\bm r},t\right) \right|^2\rangle
+
\frac{\mu_0}{2}\langle \left|{\bm B}\left({\bm r},t\right) \right|^2\rangle
\nonumber\\
=&
\int_0^\infty
d\omega \, u_{\rm tot}\left(z,\omega\right),
\end{align}
where $u_{\rm tot}\left(z,\omega\right)$ is the spectral energy density.
Assuming that the nonlocal medium is in thermal equilibrium with its surroundings and the system is rotationally invariant around the $z$ axis, this can be written in terms of the previously calculated reflection coefficients:
\begin{align}
 & \!\!\! u_{\rm tot}(z,\omega)
=
\nonumber\\
 &  \!\!\! \frac{u_0}{k_0}
\int_0^{k_0}  \!\!\!
\frac{KdK}{\sqrt{k_0^2-K^2}}
\left[
1+\frac{K^2\textrm{Re} \left[ (r_s+r_p) e^{2i\sqrt{K^2-k_0^2}|z|}  \right] }{2k_0^2}
\right]
\nonumber\\
& +
\frac{u_0}{2 k_0^3} \int_{k_0}^\infty
\frac{K^3dK}{\sqrt{K^2-k_0^2}}
\textrm{Im}[r_s + r_p]
e^{-2\sqrt{K^2-k_0^2}|z|}
.
\label{eq:utot}
\end{align}
The first integral in (\ref{eq:utot}) is the contribution of propagating waves while the second comes from evanescent waves.
The term $u_0$ is the spectral energy density in the absence of the material, given by
\begin{align}
u_0&
=
\frac{\Theta(\omega,T)\omega^2}{\pi^2c^3},
\end{align}
where the mean energy of a harmonic oscillator in thermal equilibrium is
\begin{align}
\Theta(\omega,T)&
=
\hbar\omega
\left(
\frac{1}{2}
+\frac{1}{e^{\hbar\omega/k_BT}-1}
\right).
\label{eq:theta}
\end{align}
The first term of (\ref{eq:theta}) gives rise to the electromagnetic zero-point energy. 

\begin{figure}[!htb]\centering
{\includegraphics[width=\linewidth]{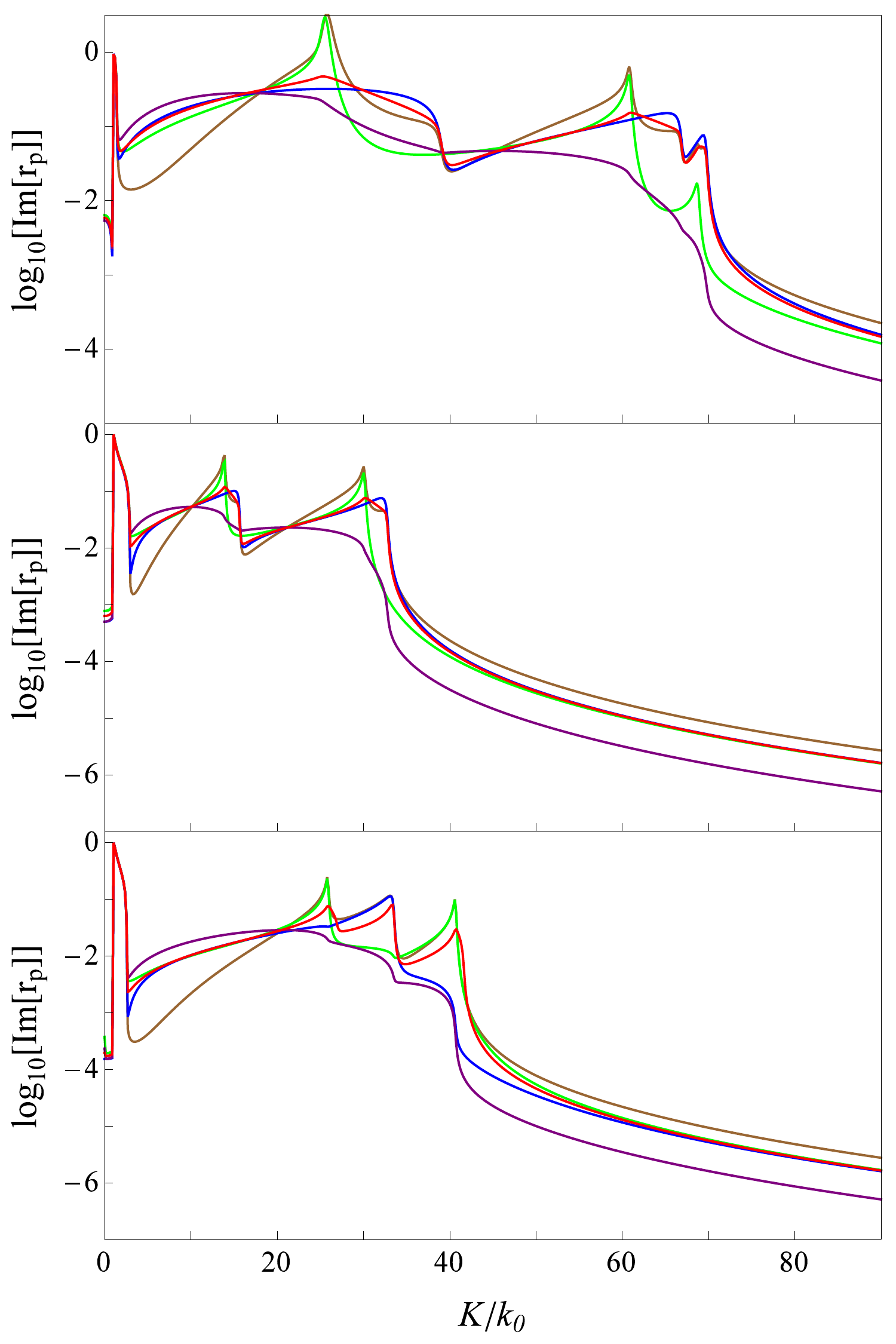}}
\caption{
Comparison of $\textrm{Im}[r_p]$ as a function of $K$ for evanescent waves in ZnO at $\hbar\omega=3.44{\rm eV}$ (top), GaAs at $\hbar\omega=1.517{\rm eV}$ (middle) and CdS ($\hat{{\bm z}}\parallel{\bm c}$) at $\hbar\omega=2.573{\rm eV}$ (bottom).
Plot styles follow the conventions in  Fig. \ref{fig:ZnO}.
}
\label{fig:Imrp}
\end{figure}

In the $K\to\infty$ limit for the local medium $r_s\to0$ and $r_p\to\chi(\omega)/(2+\chi(\omega))$, leading to the divergent result for the second integral\cite{joulain2005}:
\begin{align}
\frac{1}{4|z|^3}
\frac{\textrm{Im}[\chi(\omega)]}{|2+\chi(\omega)|^2}.
\label{eq:utot_nondispersive}
\end{align}
In our previous paper, we showed that the inclusion of spatial dispersion lead to peaks in $\textrm{Im}[r_p]$ near the point where the $\textrm{Re}[\Gamma^2]$ changed sign from positive to negative, followed by a $1/K^4$ decay in the large $K$ limit.
In Fig. \ref{fig:Imrp} we find the same behavior for ZnO, GaAs and CdS ($\hat{{\bm z}}\parallel{\bm c}$), with a peak for every $\Gamma_m$ value.

\begin{figure}[!htb]\centering
{\includegraphics[width=\linewidth]{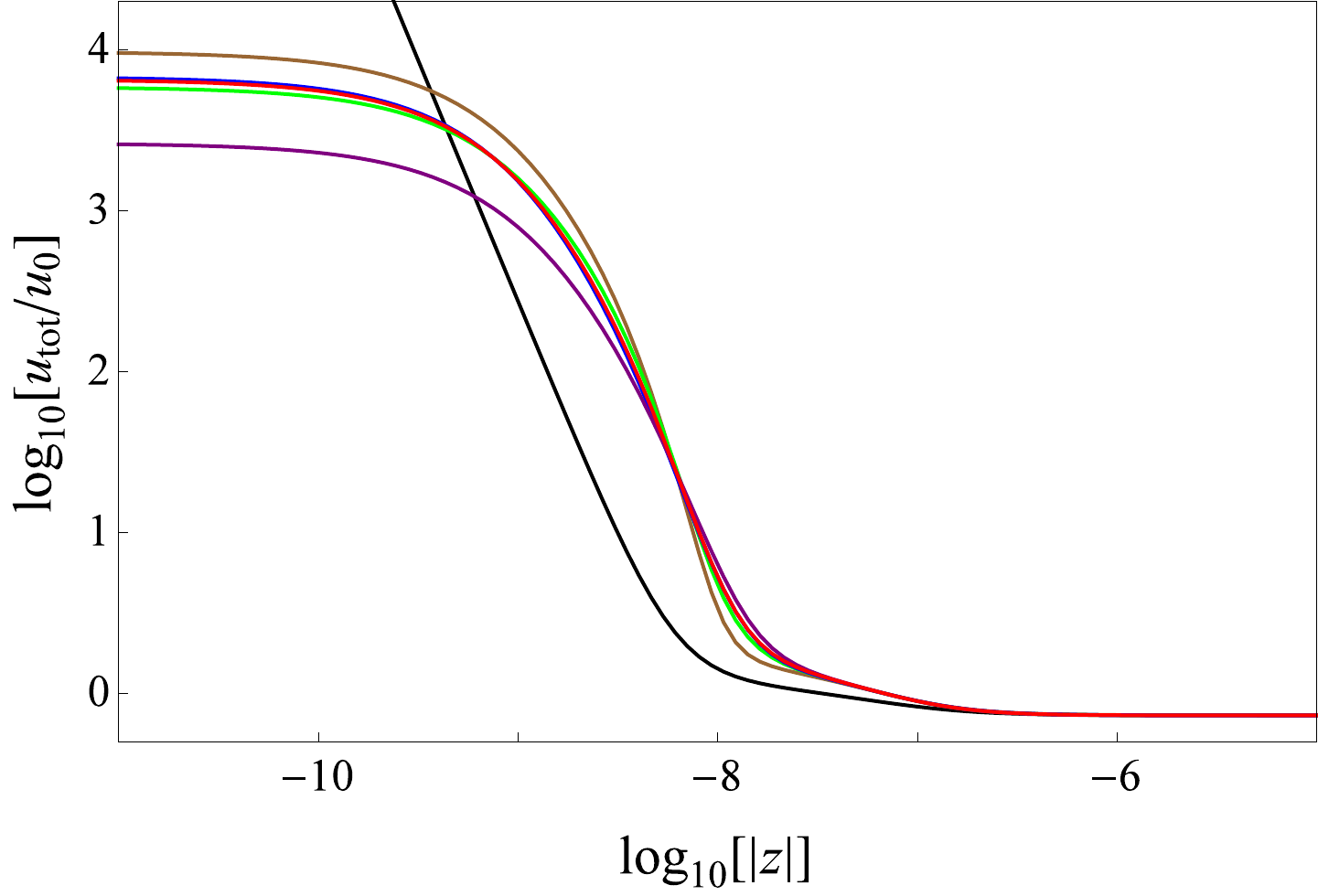}}
\caption{
Behavior of $u_{\rm tot}$ as a function of distance $|z|$ from the boundary of ZnO at $\hbar\omega=3.44{\rm eV}$ compared to the $1/z^3$ divergent result  of the local model (black line).
Plot styles follow the conventions in  Fig. \ref{fig:ZnO}.
}
\label{fig:utotz}
\end{figure}

Figure \ref{fig:utotz} shows the $z$ dependence of $u_{\rm tot}$ for ZnO at $\hbar\omega=3.44{\rm eV}$.
The behavior is the same as found in our previous paper - the $1/|z|^3$ divergence is removed and $u_{\rm tot}$ saturates to a finite value. The ABC behavior is similar, with Ting \emph{et al}. giving the largest result, Fuchs-Kleiwer, Rimbey-Mahan and Agarwal \emph{et al}. have similar intermediate values and Pekar is the smallest.
The other materials display the same behavior and so are omitted here.

\begin{figure}[!htb]\centering
{\includegraphics[width=\linewidth]{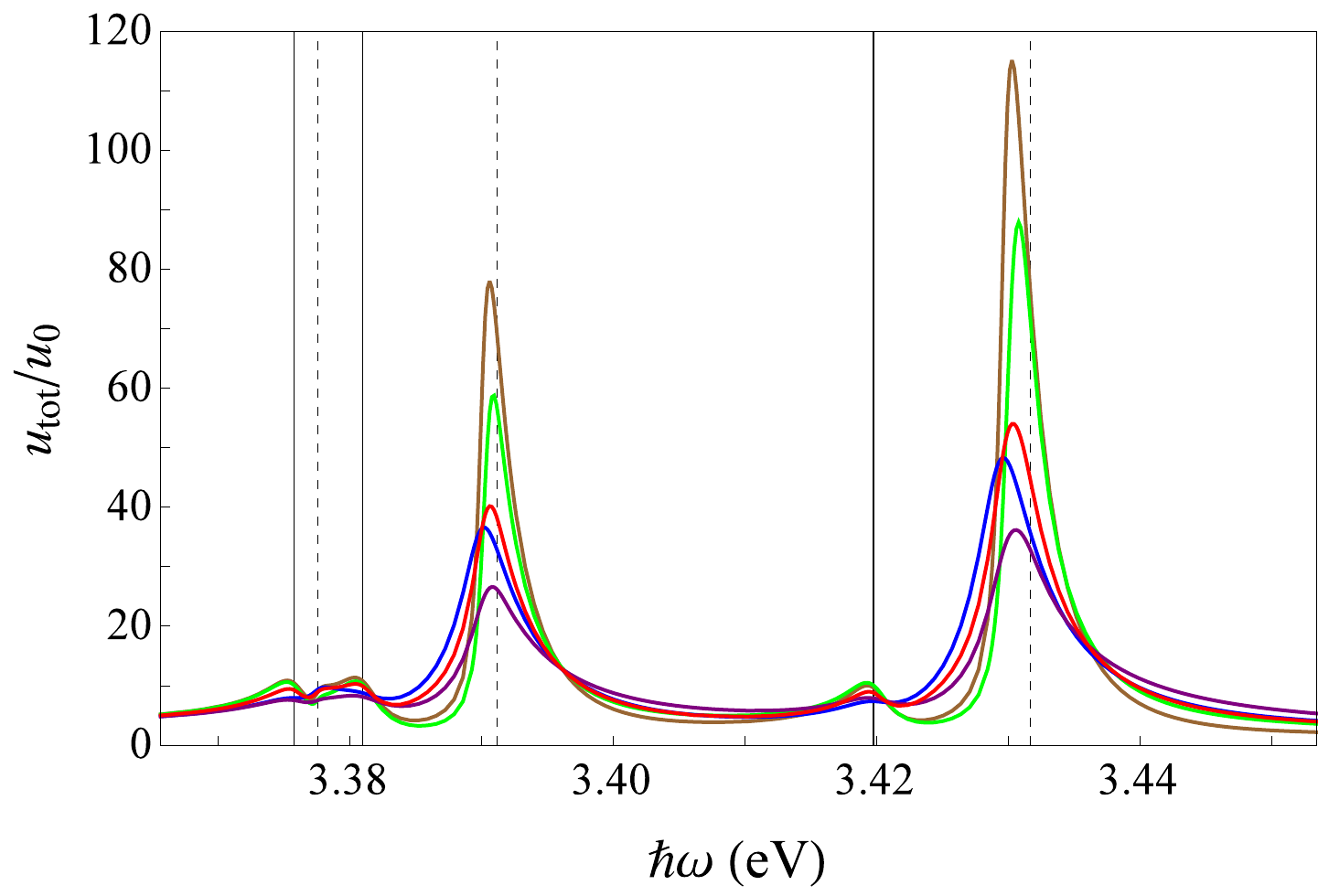}}
\caption{
Behavior of $u_{\rm tot}$ at a fixed distance of 8nm from the boundary of ZnO.
Vertical lines indicate $\omega_{Tm}$ (solid) and $\omega_{Lm}$ (dashed) values.
Plot styles follow the conventions in  Fig. \ref{fig:ZnO}.
}
\label{fig:ZnOutotw}
\end{figure}

The main differences to the one-resonance model are found in the $\omega$ dependence of $u_{\rm tot}$ at a fixed distance from the boundary.
We first consider the simple case of ZnO, with multiple, non-intersecting parabolic exciton bands.
Figure \ref{fig:ZnOutotw} shows $u_{\rm tot}(\omega)$ at a distance of 8nm from the boundary.
The results are strongly dependent on the choice of ABC, with Ting \emph{et al}. giving the largest peaks, followed by Fuchs-Kleiwer, Agarwal \emph{et al}., Rimbey-Mahan and finally Pekar.
This behavior agrees with our previous paper\cite{churchill2016b}.
Each resonance has two associated peaks in $u_{\rm tot}(\omega)$, for a total of six.
The peaks in $u_{\rm tot}$ near $\omega_{Tm}$ are due to the $s$-polarization contribution to the integral and the peaks at $\omega_{Lm}$ are due to the $p$-polarization contribution.
In the one-resonance system, the $s$-polarization peaks in $u_{\rm tot}$ were typically much smaller than their $p$-polarization counterparts.
However, the $u_{\rm tot}$ peak at $\omega_{L1}$ has been suppressed due to the proximity of the peak at $\omega_{L2}$  and is now comparable in size to the $s$-polarization peaks.
This can be seen in Fig. \ref{fig:Imrp}, where the peak in $\textrm{Im}[r_p]$ associated with $\Gamma_1$ and the $m=1$ resonance at the largest $K$ value is very small due to the presence of the nearby $\Gamma_2$ peak at a smaller $K$ value.

\begin{figure}[!htb]\centering
{\includegraphics[width=\linewidth]{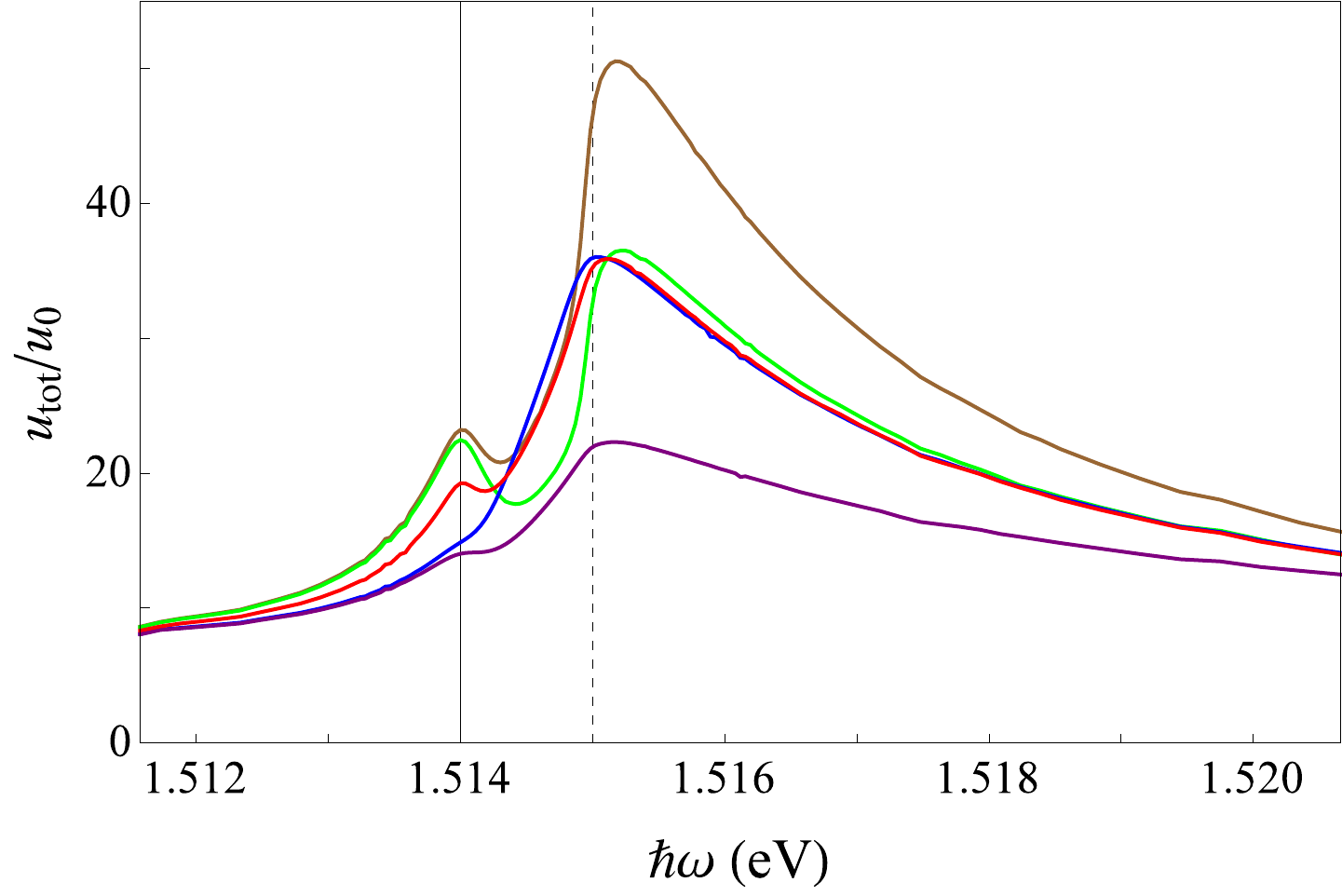}}
\caption{
Behavior of $u_{\rm tot}$ at a fixed distance of 8nm from the boundary of GaAs.
Vertical lines indicate $\omega_{T}$ (solid) and $\omega_{L}$ (dashed) values.
Plot styles follow the conventions in  Fig. \ref{fig:ZnO}.
}
\label{fig:GaAsutotw}
\end{figure}

\begin{figure}[!htb]\centering
{\includegraphics[width=\linewidth]{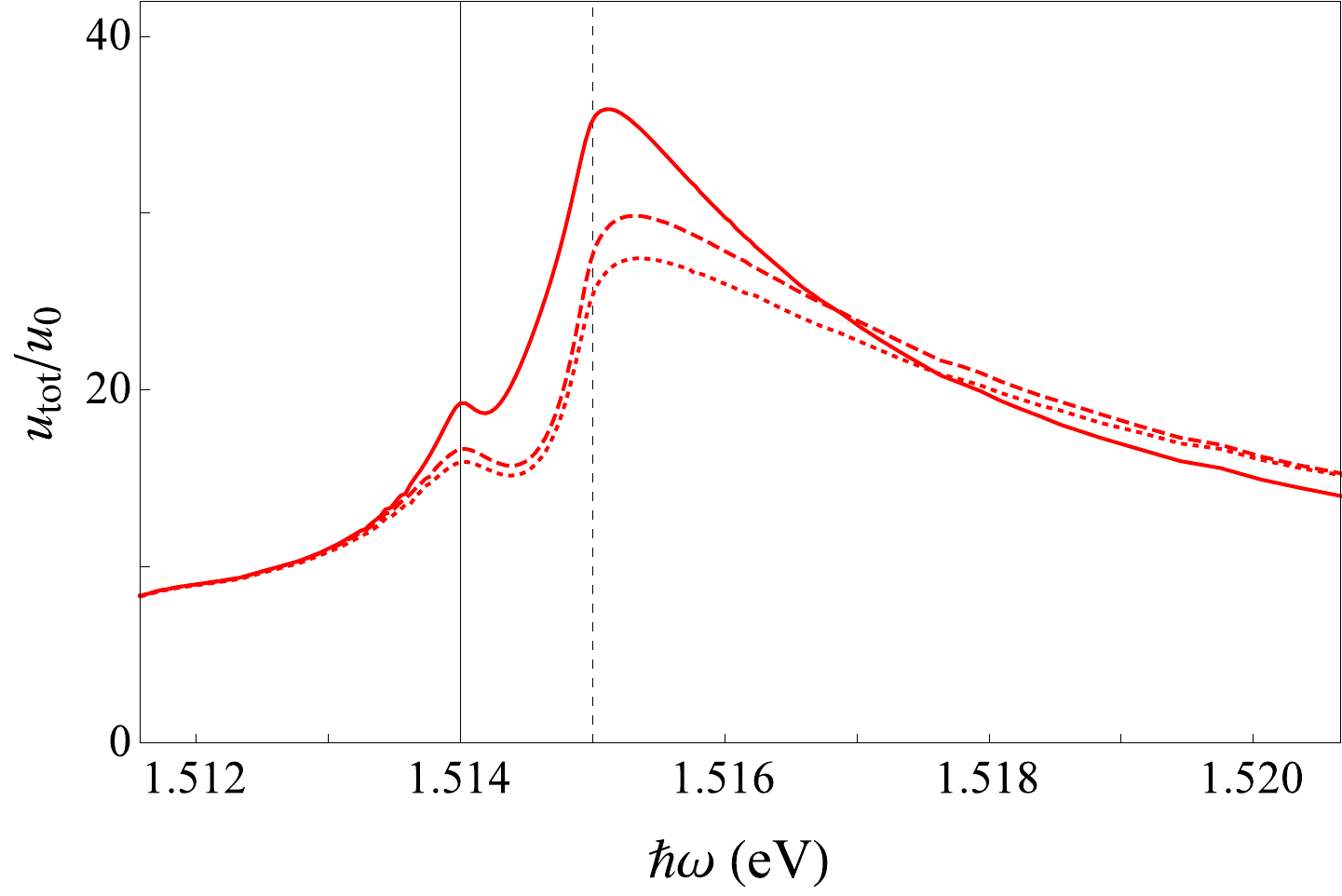}}
\caption{
Detail of the Agarwal \emph{et al}.\ ABC in Fig.\ref{fig:GaAsutotw} for the heavy/light exciton model compared to the one exciton results for $D^*$ (dashed) and $D_{\rm eff}$ (dotted).
}
\label{fig:GaAsutotw2}
\end{figure}

We next consider the heavy/light exciton model of GaAs.
Figure \ref{fig:GaAsutotw} shows $u_{\rm tot}(\omega)$ at a distance of 8nm from the boundary.
At first glance the results appear similar to those of the single resonance model, but the comparison in Fig. \ref{fig:GaAsutotw2} reveals that both the $D^*$ and $D_{\rm eff}$ single-exciton models both underestimate the peak values.
This is due to the behavior of $r_p$.
While both approximations provide a good fit for propagating waves, $u_{\rm tot}$ depends more on the evanescent wave contribution at small distances. 
Both approximations have a single peak in $\textrm{Im}[r_p]$ at large $K$ in contrast to the two peaks of the heavy/light exciton model in Fig. \ref{fig:Imrp}.

\begin{figure}[!htb]\centering
{\includegraphics[width=\linewidth]{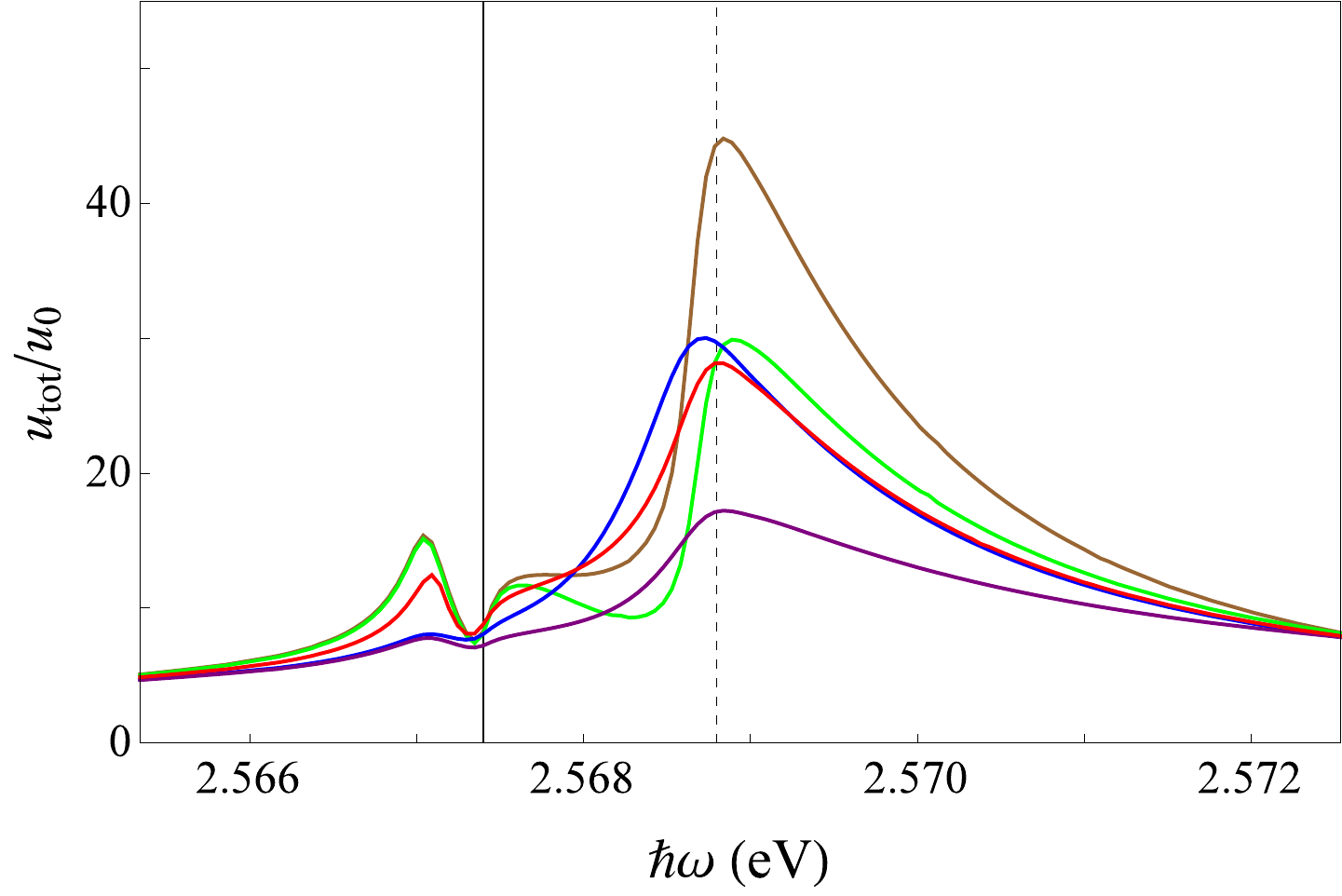}}
\caption{
Behavior of $u_{\rm tot}$ at a fixed distance of 8nm from the boundary of CdS.
Vertical lines indicate $\omega_{T}$ (solid) and $\omega_{L}$ (dashed) values.
Plot styles follow the conventions in  Fig. \ref{fig:ZnO}.
}
\label{fig:CdSutotw}
\end{figure}

We finally consider the uniaxial crystal of CdS in the ${\bm c}\parallel\hat{{\bm z}}$ case, as the other orientations lack the rotational invariance about the $z$ axis required for Eq.\ (\ref{eq:utot}).
The underlying behaviour behind Fig.\ \ref{fig:CdSutotw} is more complex than the previous cases with $k^2$ dispersion.
There are now two peaks in each of the $s$-and $p$-polarization contributions to $u_{\rm tot}$.
A large peak in the $p$-polarization term is still found near $\omega_L$, but a new smaller peak is also present just above $\omega_T$.
However, the peak in the $s$-polarization contribution to $u_{\rm tot}$ previously found at $\omega_T$ is now a minimum, with a larger peak below this frequency and a smaller peak above that coincides with the position of the lower peak in the $p$-polarization contribution.
This leads to an overall three-peak structure in $u_{\rm tot}$.
This strongly contrasts with the results in Fig.\ \ref{fig:CdS[rprs]}, where the reflection coefficients for ${\bm c}\parallel\hat{{\bm z}}$ are nearly identical to the $\xi=0$ results.
Such a contrast was also present in our previous paper\cite{churchill2016b}, where a difference in transverse and longitudinal nonlocal terms had little effect on $r_p$, but a significant effect on $u_{\rm tot}$.
The overall effect of the ABC choice remains the same as in previous sections.

From this and the previous sections, it is clear that if multiple spatially-dispersive resonances are present in a medium then they cannot be considered separately.
Each of these results displays behavior not present in the single-resonance case, such as the suppression of peaks in $u_{\rm tot}$ associated with the $p$-polarization in closely-spaced parabolic bands, the difference between the heavy/light exciton model to the single band approximations and finally the additional $u_{\rm tot}$ peak from linear splitting in exciton bands.

\section{Conclusions}

We have extended the work of Halevi and Fuchs\cite{Halevi} to derive exact expressions for electromagnetic reflection and transmission coefficients at the boundary of a medium with multiple spatially dispersive resonances in the susceptibility.
Surface effects are included by using phenomenological reflection coefficients $U_{mi}$ for the polarization waves at the boundary.
We have compared the results for several multi-resonance media, using a variety of $U_{mi}$ values corresponding to ABC's in the literature.
In the case of heavy/light exciton bands, we have found an improved fit for the single band approximation with $\sqrt{D^*}=(\sqrt{D_{\rm h}}+\sqrt{D_{\rm l}})/2$.

The model has been extended to alternate exciton dispersion relations with the inclusion of a linear splitting term in $\omega(k)$ typical of uniaxial crystals. 
We have compared the results when the crystal axis ${\bm c}$ is aligned with each of the co-ordinate axes in our system
The largest effects were seen with ${\bm c}$ perpendicular to the plane of incidence for $s$-polarization and in the plane of incidence, parallel to the surface, for $p$-polarization.

Finally we have used the calculated reflection coefficients to find the zero-point and thermal spectral energy density $u_{\rm tot}(z,\omega)$ outside the dielectric.
Many features are the same as the single-resonance model, such as the effect of the ABC choice and the saturation of $u_{\rm tot}$ as $z\to0$.
However, there is new behavior that is only present when the multiple-resonances are considered together.
We have found that close resonances can lead to significant suppression in the peaks of $u_{\rm tot}(\omega)$ and that single band approximations fail to capture the correct behavior of the heavy/light exciton band model.
The linear splitting term led to significant changes for $u_{\rm tot}$ in the uniaxial crystal by splitting the peak at the resonant frequency to give an overall three-peak structure.

While the model presented here incorporates many more of the features found in real materials than Halevi and Fuchs, it could be extended  further to include differences between the transverse and longitudinal susceptibilities\cite{churchill2016b} or  higher-order nonlocal terms.
This could also be applied to other problems, such as identifying which ABC is most appropriate for a medium with a complex exciton band structure, or in the calculation of Casimir self-forces\cite{hor14}.

\end{document}